\documentclass[%
prb
amsmath,amssymb,
reprint,%
superscriptaddress,footinbib,%
]{revtex4-2}

\usepackage{graphicx}% Include figure files
\usepackage{dcolumn}% Align table columns on decimal point
\usepackage{bm}% bold math
\usepackage[dvipsnames]{xcolor}
\usepackage{chemformula}
\usepackage{siunitx}
\usepackage{amsmath}
\usepackage{makecell}
\usepackage{multirow}

\setcitestyle{super}

\begin{document}

\title{Atomic Layer Deposition of Yttrium Iron Garnet Thin Films for 3D Magnetic Structures}
	
\author{M. Lammel}
\email{michaela.lammel@uni-konstanz.de}
\affiliation{Institute for Metallic Materials, Leibniz Institute of Solid State and Materials Science, 01069 Dresden, Germany}
\affiliation{Institute of Applied Physics, Technische Universit\"at Dresden, 01069 Dresden, Germany}
\affiliation{Fachbereich Physik, Universität Konstanz, 78457 Konstanz, Germany}

\author{D. Scheffler}
\affiliation{Institut f\"ur Festk\"orper- und Materialphysik, Technische Universit\"at Dresden, 01069 Dresden, Germany}

\author{D. Pohl}
\affiliation{Dresden Center for Nanoanalysis (DCN), cfaed, Technische Universi\"at Dresden, 01062 Dresden, Germany}

\author{P. Swekis}
\affiliation{Institut f\"ur Festk\"orper- und Materialphysik, Technische Universit\"at Dresden, 01069 Dresden, Germany}
\affiliation{Max-Planck Institute for Chemical Physics of Solids, 01187 Dresden, Germany}

\author{S. Reitzig}
\affiliation{Institut f\"ur angewandte Physik, Technische Universit\"at Dresden, 01187 Dresden, Germany}

\author{S. Piontek}
\affiliation{Institute for Metallic Materials, Leibniz Institute of Solid State and Materials Science, 01069 Dresden, Germany}

\author{H. Reichlova}
\affiliation{Institut f\"ur Festk\"orper- und Materialphysik, Technische Universit\"at Dresden, 01069 Dresden, Germany}

\author{R. Schlitz}
\affiliation{Institut f\"ur Festk\"orper- und Materialphysik, Technische Universit\"at Dresden, 01069 Dresden, Germany}
	
\author{K. Geishendorf}
\affiliation{Institute for Metallic Materials, Leibniz Institute of Solid State and Materials Science, 01069 Dresden, Germany}
\affiliation{Institute of Applied Physics, Technische Universit\"at Dresden, 01069 Dresden, Germany}
	
\author{L. Siegl}
\affiliation{Institut f\"ur Festk\"orper- und Materialphysik, Technische Universit\"at Dresden, 01069 Dresden, Germany}
\affiliation{Fachbereich Physik, Universität Konstanz, 78457 Konstanz, Germany}

\author{B. Rellinghaus}
\affiliation{Dresden Center for Nanoanalysis (DCN), cfaed, Technische Universi\"at Dresden, 01062 Dresden, Germany}

\author{L.M. Eng}
\affiliation{Institut f\"ur angewandte Physik, Technische Universit\"at Dresden, 01187 Dresden, Germany}
\affiliation{ct.qmat: Dresden-W\"urzburg Cluster of Excellence - EXC 2147, Technische Universit\"at Dresden, 01062 Dresden, Germany}

\author{K. Nielsch}
\affiliation{Institute for Metallic Materials, Leibniz Institute of Solid State and Materials Science, 01069 Dresden, Germany}
\affiliation{Institute of Applied Physics, Technische Universit\"at Dresden, 01069 Dresden, Germany}
\affiliation{ct.qmat: Dresden-W\"urzburg Cluster of Excellence - EXC 2147, Technische Universit\"at Dresden, 01062 Dresden, Germany}
\affiliation{Institute of Materials Science, Technische Universit\"at Dresden, 01069 Dresden, Germany}

\author{S.T.B. Goennenwein}
\affiliation{Institut f\"ur Festk\"orper- und Materialphysik, Technische Universit\"at Dresden, 01069 Dresden, Germany}
\affiliation{Fachbereich Physik, Universität Konstanz, 78457 Konstanz, Germany}
\affiliation{ct.qmat: Dresden-W\"urzburg Cluster of Excellence - EXC 2147, Technische Universit\"at Dresden, 01062 Dresden, Germany}
	
\author{A. Thomas}
\email{a.thomas@ifw-dresden.de}
\affiliation{Institute for Metallic Materials, Leibniz Institute of Solid State and Materials Science, 01069 Dresden, Germany}
\affiliation{Institut f\"ur Festk\"orper- und Materialphysik, Technische Universit\"at Dresden, 01069 Dresden, Germany}

\date{\today}% It is always \today, today,
	
\begin{abstract}
Magnetic nanostructures with non-trivial three dimensional shapes enable complex magnetization configurations and a wide variety of new phenomena. To date predominantly magnetic metals have been considered for non-trivial 3D nanostructures, although the magnetic and electronic transport responses are intertwined in metals. Here, we report the first successful fabrication of the magnetic insulator yttrium iron garnet (\ch{Y3Fe5O12}, YIG) via atomic layer deposition (ALD) and show that conformal coating of 3D objects is possible. We utilize a supercycle approach based on the combination of sub-nanometer thin layers of the binary systems \ch{Fe2O3} and \ch{Y2O3} in the correct atomic ratio with a subsequent annealing step for the fabrication of ALD-YIG films on \ch{Y3Al5O12} substrates. Our process is robust against typical growth-related deviations, ensuring a good reproducibility. The ALD-YIG thin films exhibit a high crystalline quality as well as magnetic properties comparable to samples obtained by other deposition techniques. The atomic layer deposition of YIG thus enables the fabrication of novel three dimensional magnetic nanostructures based on magnetic insulators. In turn, such structures open the door for the experimental verification of curvature induced changes on pure spin currents and magnon transport effects.
\end{abstract}

\maketitle

\section{Motivation}
The field of spintronics as well as commercially available spintronic devices are currently based on conventional planar (two dimensional) structures and layered stacks. In the last few years, however, significant experimental efforts went into fabricating and characterizing non-trivial, three dimensional magnetic structures. This development was promoted by the strive for new downscaling methods for data storage\cite{Parkin_2008} but it quickly became apparent that leaving the paradigm of planar thin films would lead to a variety of new physical phenomena in magnetic materials.\cite{Streubel_2016, Fernandez_Pacheco_2017,Vedmedenko_2020, Das_2019, Sheka_2020} In non-planar nanostructures such as nanotubes, for example, a curvature induced Dzyaloshinskii-Moriya interaction emerges, which does not only lead to novel magnetic configurations but is also predicted to lead to interesting emerging effects such as spin wave non-reciprocity\cite{Otalora_2016} or Cherenkov-like magnons.\cite{Yan_2013} Many of these effects have not been explored experimentally due to the difficulty to prepare structures with the necessary characteristics. One of the most promising routes to fabricate arbitrarily-shaped three dimensional magnetic structures is via coating pre-patterned objects. Atomic layer deposition (ALD) inherently gives the possibility to conformally coat surfaces of any shape, and thus is ideal for the realization of non-trivial 3D nanostructures. 
ALD is a chemical vapor deposition technique, which is based on self-terminating surface reactions. Thus, it allows for conformal coatings with a constant growth rate (growth per cycle), which enables a thickness control down to very thin layers.\cite{Miikkulainen_2013, Elers_2006} Substantial progress in precursor design within the last decade enabled the fabrication of a variety of materials by ALD. These include a multitude of ternary processes which require a careful selection and thorough understanding of the involved processes.\cite{Miikkulainen_2013,Mackus_2018} 
In fact, one of the prototypical materials used in pure spin current experiments, yttrium iron garnet (\ch{Y3Fe5O12}, YIG), is a ternary oxide. Up to date YIG thin films are typically fabricated by pulsed laser deposition\cite{Althammer_2013} or sputtering.\cite{Hauser_2017} However, both techniques are directional, which limits their usability for 3D structures.

In this letter we present a robust process for the fabrication of YIG thin films via ALD. To this end, we employ a nanolaminate approach based on the alternate depositions of \ch{Y2O3} and \ch{Fe2O3} layers of sub-nanometer thickness onto \ch{Y3Al5O12} (YAG) substrates. A subsequent annealing step initiates the interdiffusion of the individual layers and promotes the crystallization into YIG. We use X-ray diffraction (XRD) and high resolution transmission electron microscopy (HR-TEM) to demonstrate a high crystalline quality of the ALD-YIG. Also, the influence of the annealing temperature, the stoichiometry and the individual layer thicknesses is examined, confirming the robustness of the nanolaminate approach. Characterizing the static and dynamic magnetic properties by SQUID magnetometry and broadband ferromagnetic resonance (bb-FMR) measurements, respectively, shows a magnetic behavior comparable to samples fabricated by pulsed laser deposition or sputtering. Finally, we validate the conformity of the deposition, showing that our ALD process indeed enables the coating of three dimensional objects. Our work thus facilitates using YIG, one of the prototypical materials in the field of insulator spintronics, also in non-planar and curved structures.

\section{Deposition of ALD-YIG thin films}
\subsection{The supercycle approach}
For the atomic layer deposition of ternary metal oxides, such as YIG, different approaches have been proposed.\cite{Mackus_2018} For the ALD of YIG, we here use the supercycle approach, which is based on the combination of two binary ALD processes (i.e. processes with two elements) in the correct stoichiometry.\cite{Mackus_2018} Typically, a post-deposition annealing step is used to achieve a homogeneous intermixing of the two individual layers and to promote crystallization.\cite{Mackus_2018} Despite from the growing list of ternary materials realized by ALD, the fabrication of YIG via ALD has not been reported so far. 

\begin{figure}[htbp]%
	\includegraphics[width=\linewidth]{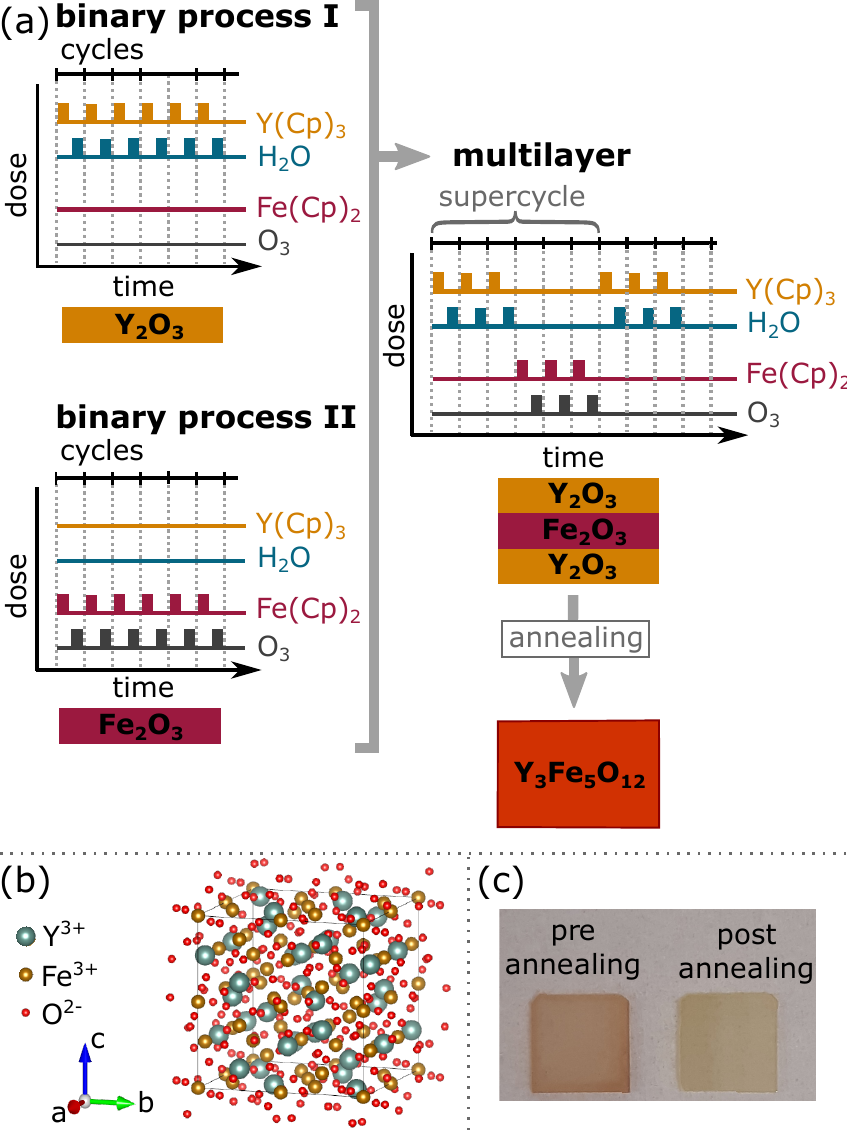}%
	\caption{A schematic of combining two binary ALD processes via the supercyle approach for the deposition of a ternary compound is given in panel (a). Binary process I describes the deposition of \ch{Y2O3} in a cyclic manner, binary process II, on the other hand, the one of \ch{Fe2O3} in the same way. By alternately depositing thin layers of the two materials, a multilayer is achieved. In panel (b) the crystal structure of YIG is depicted. Panel (c) shows the color change observed in ALD deposited nanolaminates upon heat treating them at \SI{700}{\degreeCelsius} for \SI{4}{\hour} in \SI{3}{\milli\bar} \ch{O2}.}%
	\label{ALD_YIG:nanolaminate_schematic}%
\end{figure}

Well established binary ALD processes are key for the successful fabrication of a ternary compound by the nanolaminate approach. For the deposition of YIG (\ch{Y3Fe5O12}) a combination of Y$_\mathrm{x}$O$_\mathrm{y}$ and Fe$_\mathrm{x}$O$_\mathrm{y}$ needs to be employed.
A schematic of the supercycle approach is depicted in Fig.~\ref{ALD_YIG:nanolaminate_schematic}.
Within binary process I, tris(cyclopentadienyl)yttrium (\ch{Y(Cp)3}) is used as metal-organic precursor and \ch{H2O} as co-reactant. Using them in the ALD-typical cyclic manner results in a layer of \ch{Y2O3}. Likewise, binary process II uses bi(cyclopentadienyl)iron (ferrocene, \ch{Fe(Cp)2}) as metal-organic precursor and \ch{O3} as co-reactant to deposit a layer of \ch{Fe2O3}. 
Alternately using these two binary processes results in a multilayer stack. Each so called supercycle consists of a specific number of ALD cycles of binary processes I and II. Due to the fact that usually the individual layers are of (sub-)\si{\nano\meter} thickness, the resulting multilayer stack consisting of several supercycles is also denoted as nanolaminate. By changing the number of cycles of process I and II, the thickness of the individual layers within each supercycle can be adjusted. Therefore, by changing the total number of cycles within the supercycles as well as the ratio of binary process I to binary process II, the initial intermixing and stoichiometry of the nanolaminate can be elegantly tuned. 

\subsection{Determination of the correct composition within the nanolaminates}
To ensure the correct composition within the nanolaminate, the thicknesses of the individual layers within one supercycle have been chosen to represent the atomic ratio of the metals in the final YIG layer, i.e. $\ch{Y}:\ch{Fe}=3:5$. 
This atomic ratio can be related to the thicknesses of the individual layers by\cite{Askeland_1996}
\begin{equation}
	\frac{d_{\ch{Y2O3}}\cdot\rho_{\ch{Y2O3}}}{M_{\ch{Y2O3}}} = \frac{3}{5}\cdot \frac{d_{\ch{Fe2O3}}\cdot\rho_{\ch{Fe2O3}}}{M_{\ch{Fe2O3}}},
	\label{Eq:atomic_ratio_thickness}%
\end{equation}
where the thicknesses of the corresponding layers are given by $d_{\ch{Y2O3}}$ and $d_{\ch{Fe2O3}}$ and the mass densities of \ch{Y2O3}\cite{Villars_2016_Y2O3} and \ch{Fe2O3}\cite{Landolt-Bornstein:hematite_2000} are $\rho_{\ch{Y2O3}}$$=\SI{5.01}{\gram\per\cubic\centi\meter}$ and $\rho_{\ch{Fe2O3}}$$=\SI{5.24}{\gram\per\cubic\centi\meter}$, respectively. The molar masses are $M_{\ch{Y2O3}}=\SI{225.81}{\gram\per\mole}$ and $M_{\ch{Fe2O3}}=\SI{159.69}{\gram\per\mole}$. 
One layer of Y-O was taken as smallest building block for the nanolaminates. To that end the thickness of one Y-O monolayer was estimated to be \SI{0.3}{\nano\meter} by considering the sum of typical sizes of the individual atoms.\cite{Clementi_1963, Slater_1964} Hence, the thicknesses of the \ch{Y2O3} and \ch{Fe2O3} layers can be related while simultaneously ensuring the correct stoichiometry by
\begin{equation}
	\begin{split}
		d_{\ch{Y2O3}} &= \SI{0.3}{\nano\meter}\\
		d_{\ch{Fe2O3}} &= C^{-1} \cdot \SI{0.3}{\nano\meter}\\
		\text{with } C &=\frac{3}{5}\cdot \frac{\rho_{\ch{Fe2O3}}}{M_{\ch{Fe2O3}}} \cdot \frac{M_{\ch{Y2O3}}}{\rho_{\ch{Y2O3}}}.
	\end{split}
	\label{Eq:coef_thickness}%
\end{equation}
Taking into consideration the growth rates (growth per cycle, gpc) of the \ch{Fe2O3} and \ch{Y2O3} layers, i.e. gpc$(\ch{Fe2O3})=\SI{0.016}{\nano\meter}$/cycle and gpc$(\ch{Y2O3})=\SI{0.092}{\nano\meter}$/cycle, the number of cycles can be determined for a predefined thickness of the individual layers by $\# \mathrm{cycles} = d/\mathrm{gpc}$. Hereby, the accurate determination of the "in-situ" growth rates of \ch{Fe2O3} on \ch{Y2O3} and vice versa is crucial for obtaining the desired stoichiometry in the nanolaminates. "In-situ" refers to the fact that the direct deposition of layers with different chemical compositions on top of each other without opening the ALD reactor can take place at qualitatively different rates as compared to "ex-situ" deposition. The latter corresponds to the growth rate upon exposing the sample to ambient atmosphere in between layers. It has been reported that for "in-situ" growth etching processes can occur for the subsequent ALD growth of different layers due to the reactions of the different precursors with the previous layer.\cite{Wiegand_2018,Myers_2021} To account for such influences, we determined the "in-situ" as well as the "ex-situ" growth rates of the \ch{Fe2O3} and \ch{Y2O3} processes carefully by using X-ray fluorescence.\cite{Lammel_2020,Lammel_PhD} By combining the relation between growth rate and layer thickness with Eq. (\ref{Eq:coef_thickness}), the number of cycles needed for one building block of the nanolaminate with \SI{0.3}{\nano\meter} while maintaining the proper atomic ratio was calculated by
\begin{equation}
	\begin{split}
		\#\,\mathrm{cycles}(\ch{Y2O3}) &= \frac{\SI{0.3}{\nano\meter}}{\mathrm{gpc}(\ch{Y2O3})}\\
		\#\,\mathrm{cycles}(\ch{Fe2O3}) &= C^{-1} \cdot \frac{\SI{0.3}{\nano\meter}}{\mathrm{gpc}(\ch{Fe2O3})},
	\end{split}
	\label{Eq:cycles_complete}%
\end{equation}
which results in $\#\,\mathrm{cycles}(\ch{Y2O3}) = 4$ and $\#\,\mathrm{cycles}(\ch{Fe2O3}) = 21$. Using these cycle numbers, one \ch{Y2O3}/\ch{Fe2O3} double layer, i.e. one supercycle, features a thickness of approximately two monolayers. With this, the number of sets needed for the deposition of a YIG nanolaminate with a specific thickness $d_{\ch{YIG}}$ can be determined by 
\begin{equation}
	\#\,\mathrm{sets} = \frac{d_{\ch{YIG}}}{d_{\ch{Y2O3}}+d_{\ch{Fe2O3}}}.
	\label{Eq:thickness_laminate}% 
\end{equation}
For a YIG thickness of $d_{\ch{YIG}}=\SI{30}{\nano\meter}$ and an individual layer thickness of one monolayer, this results in 47 sets. We chose a YIG layer thickness of $d_{\ch{YIG}}=\SI{30}{\nano\meter}$ for it allows for easy characterization, especially by SQUID magnetometry and X-ray diffraction, while simultaneously keeping the process time reasonably short.

The nanolaminates were deposited in a Gemstar-XT thermal ALD reactor from Arradiance, which was operated in stop-exposure mode. The metal-organic precursors, i.e. \ch{Fe(Cp)2} and \ch{Y(Cp)3}, were supplied by Strem. The ozone was generated by a BMT 803N ozone generator from pure oxygen.
Single crystalline YAG (\ch{Y3Al5O12}) wafer pieces from CrysTec with an out of plane orientation of (111) were used as purchased as substrates for the fabrication of the nanolaminates, due to their lattice mismatch of \SI{3.13}{\percent} with YIG and their diamagnetic nature.

To account for a considerable growth delay of the first layer on the substrate, the deposition of the nanolaminates was started with an \ch{Y2O3} buffer layer of 33 cycles. For the subsequent nanolaminate the buffer layer deposition provides a more complete activation of the surface-active groups. The growth delay of the in-situ deposited layers has been studied extensively in previous publications.\cite{Lammel_2020, Lammel_PhD}
A \ch{Y2O3} cycle number of 4 corresponds to a calculated thickness of one atomic layer. The number of \ch{Fe2O3} cycles was chosen accordingly to result in the ideal $\ch{Y}:\ch{Fe}$ ratio of $3:5=0.6$. Using 47 supercycles then results in a nominal nanolaminate thickness of \SI{30}{\nano\meter}. For straightforward identification the process specifics are denoted in the following by (x/y/z), with x and y the number of \ch{Y2O3} and \ch{Fe2O3} cycles, respectively, and z the number of supercycles.
The growth parameters used for the deposition as well as the number of cycles and sets for the deposition of an approximately \SI{30}{\nano\meter} thick nanolaminate are collected in Tab.~\ref{Tab:ALDYIG_growth_parameters}.
	\begin{table}[htb]
		\centering
		\begin{tabular}{ c c c c c}
			\multicolumn{5}{c}{\textbf{\ch{Y2O3}/\ch{Fe2O3} nanolaminate}}\\ 
			\hline
			\hline
			&$T_{\mathrm{chamber}}$ (\si{\degreeCelsius})& 220 &  &\\
			&$T_{\ch{Y(Cp)3}}$ (\si{\degreeCelsius}) & 150 & &\\
			&$T_{\ch{Fe(Cp)2}}$ (\si{\degreeCelsius})  & 80 & &\\ 
			\hline
			\textbf{buffer layer}&&$t_{\mathrm{p}}/t_{\mathrm{exp}}/t_{\mathrm{pump}}$ (\SI{}{\second})& \makecell{\textbf{cycles}}& \makecell{\textbf{sets}}\\
			\multirow{2}{*}{\ch{Y2O3}} & \ch{Y(Cp)3} & 0.8$_{\ch{N2}}$/1/15 & \multirow{2}{*}{33} & \multirow{2}{*}{1}\\  
			& \ch{H2O} & 0.05/1/15 & &\\   
			\hline
			\textbf{nanolaminate} & & $t_{\mathrm{p}}/t_{\mathrm{exp}}/t_{\mathrm{pump}}$ (\SI{}{\second})&\makecell{\textbf{cycles}}& \makecell{\textbf{sets}}\\
			\multirow{2}{*}{\ch{Y2O3}} & \ch{Y(Cp)3} & 0.8$_{\ch{N2}}$/1/15 & \multirow{2}{*}{4} & \multirow{5}{*}{47}\\  
			& \ch{H2O} & 0.05/1/15 &&\\   
			&&&&\\
			\multirow{2}{*}{\ch{Fe2O3}} & \ch{Fe(Cp)2} & 2/2/15 &\multirow{2}{*}{21}&\\  
			& \ch{O3} & 1/2/15 &&\\ 
		\end{tabular}
		\caption{Process parameters for the atomic layer deposition of \ch{YIG} by depositing \ch{Y2O3}/\ch{Fe2O3} nanolaminates for an individual layer thickness of one monolayer and a nominal total layer thickness of \SI{30}{\nano\meter}. For the \ch{Y2O3} deposition a \SI{5}{\milli\second} \ch{N2} inject was used prior to pulsing the metal-organic precursor to enhance the amount of precursor in the chamber. The oxidizers for both binary processes have been kept at room temperature.}
		\label{Tab:ALDYIG_growth_parameters}%
	\end{table}
To induce a complete intermixing of the individual nanolaminate layers and to promote the crystallization into YIG, several as-deposited nanolaminates have been annealed at temperatures between \SI{600}{\degreeCelsius} and \SI{1000}{\degreeCelsius} for \SI{4}{\hour}. The temperature window was chosen according to previous studies, which report that temperatures above \SI{600}{\degreeCelsius} are necessary for a successful crystallization.\cite{Kumar_2008,Park_2001,Yamamoto_2004} The annealing was performed in a reduced oxygen atmosphere of \SI{3}{\milli\bar} to counteract a possible oxygen loss in the sample at elevated temperatures.\cite{Ma_2017}

The crystal structure of YIG is schematically depicted in panel (a) of Fig.~\ref{ALD_YIG:nanolaminate_schematic}. On the left in pancel (c) of Fig.~\ref{ALD_YIG:nanolaminate_schematic} a YIG-nanolaminate grown on a YAG substrate before the annealing procedure is shown. The brown-red color of the sample stems from the \ch{Fe2O3} layers. However, upon annealing the sample clearly changes to a yellow-green color (cp. Fig.~\ref{ALD_YIG:nanolaminate_schematic}, panel (c)), which already indicates the formation of YIG.\cite{Ferrand_1972} 
In the following we will demonstrate that our ALD process indeed yields high quality YIG films.

\section{Characterization of ALD-YIG thin films}
\subsection{Structural characterization}
To validate that the nanolaminates indeed crystallize into YIG, X-ray diffraction measurements have been performed on a Bruker D8 advance with a Co anode. 
Since we refrained from using a Fe filter (which induces step-like features in the XRD spectra around \SI{60}{\degree}), diffraction peaks both from the $\ch{Co}K_{\alpha,1}$ and from the $\ch{Co}K_{\alpha,2}$ wavelength must be considered. While these individual peaks can be resolved for the YAG substrate diffraction peak, the two contributions cannot be distinguished for any of the YIG reflexes. For simplicity's sake, only the nominal position stemming from the $K_{\alpha,1}$ wavelength is considered in the analysis. 

\begin{figure*}[htbp]%
	\includegraphics{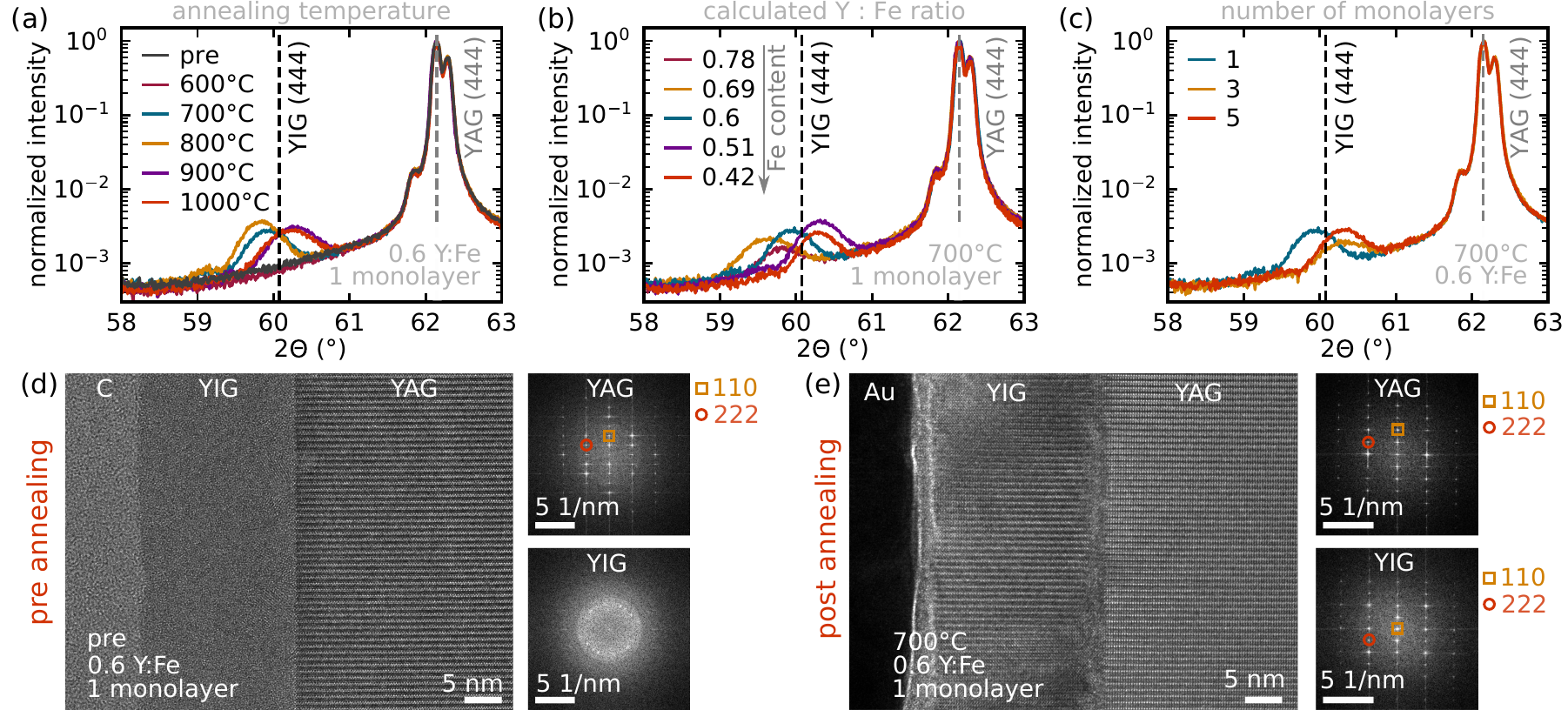}%
	\caption{X-ray diffraction spectra showing that for annealing temperatures higher than \SI{600}{\degreeCelsius}, the nanolaminates are converted into crystalline YIG layers (panel (a)). Panel (b) shows XRD measurements of nanolaminate samples featuring different nominal $\ch{Y}:\ch{Fe}$ ratios. The values are up to $\pm\SI{30}{\percent}$ off from the optimal ratio of $3:5=0.6$. The light grey arrow denotes the increase of \ch{Fe} content for decreasing atomic ratio. XRD measurements on samples with different nominal thickness of the \ch{Y2O3} layer (1 to 5 monolayers) are depicted in panel (c). The thickness of the \ch{Fe2O3} layer was adapted accordingly to yield an atomic ratio of $0.6$. The samples of all series were annealed for \SI{4}{\hour} under an oxygen pressure of \SI{3}{\milli\bar}. Panel (d) shows a cross-sectional TEM image of an ALD-YIG layer on a YAG substrate as well as the FFTs obtained from the image prior to the annealing step. The cross-sectional TEM image and the FFTs of the YAG substrate and the ALD-YIG layer after the annealing are shown in panel (e).}%
	\label{ALD_YIG:XRD_TEM}%
\end{figure*}

Panel (a) of Fig.~\ref{ALD_YIG:XRD_TEM} shows XRD spectra of samples fabricated via a supercycle process (4/21/47) and a thickness of \SI{22\pm2}{\nano\meter} for different annealing temperatures.
No reflection that can be assigned to YIG is observed in the XRD spectrum of the as-deposited (pre) samples (cp. Fig.\ref{ALD_YIG:XRD_TEM}, panel (a)). This indicates that the thermal energy during the ALD process is insufficient for promoting crystallization. Similarly, no crystallization into YIG is observed for the sample annealed at \SI{600}{\degreeCelsius}, which is in line with annealing studies for other deposition techniques.\cite{Kumar_2008,Cao_Van_2018}
A clear diffraction maximum is evolving close to the nominal position of the YIG (444) reflection,\cite{Gates_Rector_2019_YIG} i.e. $2\Theta=\SI{60.07}{\degree}$, for annealing the sample at temperatures above \SI{600}{\degreeCelsius}.
For temperatures of \SI{900}{\degreeCelsius} and higher, the maximum intensity of the YIG reflection shifts to higher $2\Theta$ values, indicating a reduction of the (out-of-plane) lattice constant. This could stem from oxygen loss in the sample or an increased interdiffusion of \ch{Al} from the YAG substrate into the YIG layer.\cite{Musa_2017}
However, the influence of both would be expected to increase with increasing temperature. This is not seen from our experimental data, where the YIG (444) reflex of the sample annealed at \SI{900}{\degreeCelsius} and the one annealed at \SI{1000}{\degreeCelsius} show the same behavior. For the samples annealed at \SI{700}{\degreeCelsius} and \SI{800}{\degreeCelsius}, the peak positions are shifted to lower $2\Theta$ values with respect to the nominal YIG (444) reflex, implying an increase in lattice constant.
Taken together, the data confirm the formation of crystalline YIG for annealing the nanolaminates at temperatures of \SI{700}{\degreeCelsius} and above.
Additionally, we also have carried out more systematic high-resolution X-ray diffraction (HR-XRD) measurements on the sample annealed at \SI{800}{\degreeCelsius} (not shown). The rocking curve of the YIG (444) reflex has a FWHM of \SI{0.7}{\degree}. This is of the same order of magnitude as values reported for high quality YIG films fabricated by pulsed laser deposition on YAG substrates (FWHM $\approx\SI{0.1}{\degree}$).\cite{Althammer_2013} The average size of the out of plane crystallites can be assessed by the Scherrer equation.\cite{Patterson_1939} Using the FWHM as well as the maximum peak position extracted from the $2\theta$-$\omega$ scans, the average crystallite size is \SI{20}{\nano\meter}, which is of the same order of magnitude as the anticipated film thickness indicating a good crystallization.

To account for deviations in stoichiometry due to uncertainties in the determination of the growth rate,\cite{Lammel_2020} we have also fabricated samples with a stoichiometry deviating by up to $\pm\SI{30}{\percent}$ from the nominal atomic ratio of $\ch{Y}:\ch{Fe}=3:5=0.6$, i.e. between $0.42<\ch{Y}:\ch{Fe}<0.78$. 
A lower atomic ratio corresponds to a higher content of Fe with respect to Y, and vice versa. For an atomic ratio of $0.78$ the process specifics were changed to (4/16/47) and for $0.42$ to (4/30/47).
The \ch{Fe2O3} cycle number has been varied rather than the \ch{Y2O3} one, due to its lower growth rate, which allows for a more refined adjustment of the stoichiometry. Panel (b) of Fig.~\ref{ALD_YIG:XRD_TEM} shows the evolution of the X-ray diffraction pattern with changing atomic ratio. All samples were annealed at \SI{700}{\degreeCelsius} for \SI{4}{\hour} in \SI{3}{\milli\bar} oxygen atmosphere.
For an atomic ratio smaller than $0.6$, the diffraction maxima reside at higher $2\Theta$ angles with respect to the nominal YIG (444) reflex, indicating a lower out-of-plane (oop) lattice constant for a nominally higher Fe content. In contrast, increasing the atomic ratio of $\ch{Y}:\ch{Fe}$ leads to a shift of the XRD peak position towards lower values, i.e. to larger oop lattice constants. The change in lattice constant is of the same order of magnitude as has been reported previously for pulsed laser deposited YIG thin films of the same nominal composition, where also a post-deposition annealing step was employed.\cite{Onbasli_2014} Despite varying the stoichiometry by $\pm\SI{30}{\percent}$, only the YIG (444) reflex is observed for all compositions, suggesting the absence of other phases. This also demonstrates that our nanolaminate approach is highly stable and will lead to the deposition of YIG even for deviations of the nominal atomic ratio of up to $\pm\SI{30}{\percent}$, which simplifies the reproducible fabrication of ALD-YIG thin films.
To further confirm the stoichiometry, commercial inductively coupled plasma optical emission spectroscopy was performed on two optimized samples with a nominal Fe:Y ratio of 5:3 (= 1.7:1) but different individual layer thicknesses. With a measured ratio of Fe:Y = 2.0(1):1 they showed a good agreement with the expected value, confirming the reproducibility of the atomic ratio for an increasing individual layer thickness.

Apart from changing the stoichiometry, also the number of monolayers within the two binary processes have been varied. By increasing the individual layer thicknesses, influences of nonlinearities in the initial growth regime are mitigated due to the decreased number of interfaces. However, thicker individual layers might result in an incomplete interdiffusion during the annealing step. The XRD spectra for samples with a calculated \ch{Y2O3} layer thickness of 1, 3 and 5 monolayers, are given in panel (c) of Fig.~\ref{ALD_YIG:XRD_TEM}. The number of \ch{Fe2O3} cycles was adjusted accordingly to result in an atomic ratio of $0.6$ and the total number of supercycles to yield the same nanolaminate thickness for all combinations. This leads to process specifics of (4/21/47), (12/62/16) and (20/104/9) for assuming 1, 3 and 5 monolayers. A post-annealing step at \SI{700}{\degreeCelsius} for \SI{4}{\hour} in \SI{3}{\milli\bar} oxygen has been applied to all samples. A shift towards higher $2\Theta$ angles is evident for the deposition of more than one monolayer per supercycle, which is comparable to the shifts observed for changing the temperature and atomic ratio. However, no delay of the crystallization due to a larger individual layer thickness is observed.

To obtain a deeper insight into the crystalline quality of our ALD-YIG layers, high resolution transmission electron microscopy (HR-TEM) images of as-deposited ALD-YIG are compared to the ones of annealed samples. 
For the HR-TEM measurements, lamellae of the samples were prepared by means of a focused gallium ion beam (FIB) of a Helios NanoLab 600i from FEI. The samples were coated with carbon and gold (as-deposited sample) or purely with gold (annealed sample) prior to the FIB cutting. This prevents an amorphisation by the ion beam during the preparation due to charging effects of the otherwise isolating samples. For the acquisition of the HR-TEM images, a JEOL JEM F200 (scanning) transmission electron microscope equipped with a cold field emission gun was used at an operation voltage of $\SI{200}{\kilo\volt}$.
The cross-sectional HR-TEM image of a ALD-YIG layer on a YAG substrate is given in Fig.~\ref{ALD_YIG:XRD_TEM}, panel (d). The YAG substrate shows the expected perfect crystallinity, whereas the as-deposited YIG layer is predominantly amorphous with only small crystals close to the YAG interface. The fast Fourier transforms (FFT) clearly support these observations. No \ch{Y2O3}/\ch{Fe2O3} layer structure is observed in the ALD-YIG layer by locals EDS analysis, which suggests that the numbers of cycles of the two binary processes are too low to result in the formation of closed layers. TEM indexing of the reflections in the FFTs is given next to the respective picture. Panel (e) of Fig.~\ref{ALD_YIG:XRD_TEM} shows the HR-TEM image after annealing the sample at \SI{700}{\degreeCelsius} for \SI{4}{\hour} in \SI{3}{\milli\bar} oxygen. The cross-sectional HR-TEM and FFT of the YIG layer demonstrate a high degree of crystallinity of the ALD-YIG layer after the heat treatment. The TEM indexing is again given next to the respective pictures. From the extracted lattice plane distances of YAG and YIG ($d_{\mathrm{YAG},110} = \SI{8.39951}{\angstrom}$, $d_{\mathrm{YAG},222} = \SI{3.45459}{\angstrom}$, $d_{\mathrm{YIG},110} = \SI{9.06793}{\angstrom}$ and $d_{\mathrm{YIG},222} = \SI{3.60585}{\angstrom}$) the lattice constant $a$ of YAG and YIG can be calculated by $a=d\sqrt{h^2+k^2+l^2}$ assuming a cubic lattice. The average lattice constants $a_{\mathrm{YAG}}=\SI{1.19\pm 0.01}{\nano\meter}$ and $a_{\mathrm{YIG}}=\SI{1.27\pm0.04}{\nano\meter}$ fit very well to the literature values of $a_{\mathrm{YAG,lit}}$\cite{CrysTec_YAG}$=\SI{1.20}{\nano\meter}$ and $a_{\mathrm{YIG, lit}}$\cite{Gilleo_1958}$=\SI{1.24}{\nano\meter}$. This further corroborates that crystalline YIG is formed during the annealing step. The slight mismatch indicates the presence of a thin strained (interfacial) layer. At the interface, a defect-rich transition region occurs due to the lattice mismatch between YAG and YIG or a surplus of yttrium from the buffer layer. An amorphous layer of approximately \SI{3}{\nano\meter} is found at the sample surface, which might result from a change in chemical composition due to an increased oxygen loss at the surface or an incomplete crystallization due to an insufficient annealing time.

Overall, the structural characterization demonstrates that high quality crystalline YIG is formed during the annealing step. 

\subsection{Magnetic characterization}
\begin{figure}[htbp]%
	\includegraphics{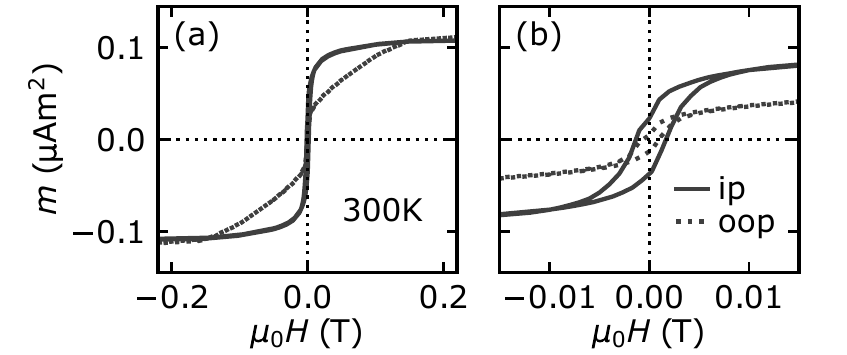}%
	\caption{The evolution of the magnetic moment of the sample annealed at \SI{800}{\degreeCelsius} for the external magnetic field applied in-plane (ip) and out of the sample plane (oop) is given in panel (a). To reveal the coercive fields, panel (b) shows the same data around $\mu_0 H =\SI{0}{\tesla}$.}%
	\label{ALD_YIG:squid}%
\end{figure}

From the perspective of spintronics, two of the most desired characteristics of YIG thin films are a small coercive field as well as a low magnetization damping (narrow ferromagnetic resonance linewidth). To determine the static magnetic properties of our ALD-YIG samples, we used SQUID magnetometry measurements using a MPMS-XL7 from Quantum Design with reciprocrating sample option. To keep parasitic moments as low as possible, the samples were mounted exclusively in-between two or more straws provided by Quantum Design. The change in magnetic moment with the external magnetic field was recorded for fields between $\pm\SI{1}{\tesla}$ at a resolution of \SI{1}{\milli\tesla}.
The best structural qualities were found for annealing the samples at \SI{700}{\degreeCelsius} and \SI{800}{\degreeCelsius} for \SI{4}{\hour} in \SI{3}{\milli\bar} \ch{O2} (cp. Fig.~\ref{ALD_YIG:XRD_TEM}, panel (a)). As the the sample annealed at \SI{700}{\degreeCelsius} was used for the TEM investigation, the magnetic characterization was performed on the sample annealed at \SI{800}{\degreeCelsius}. In Fig.~\ref{ALD_YIG:squid}, the magnetic moment of this sample as a function of the external magnetic field at \SI{300}{\kelvin} is shown for different field ranges. The linear background stemming from the diamagnetic YAG substrate was subtracted from the data.
At \SI{300}{\kelvin}, an external magnetic field of \SI{0.05}{\tesla} and \SI{0.15}{\tesla} is needed for saturating the moment at $m_{\mathrm{sat}}=\SI{0.11}{\micro\ampere\meter\squared}$ for the magnetic field applied in and perpendicular to the film plane, respectively. The coercive field for an external field applied in the film plane is $\mu_0H_{\mathrm{c}}=\SI{1.5}{\milli\tesla}$, which is lower than that of high-quality films fabricated by pulsed laser deposition ($\mu_0H_{\mathrm{c}}=\SI{5}{\milli\tesla}$)\cite{Althammer_2013} or post-annealed sputtered samples ($\mu_0H_{\mathrm{c}}\geq\SI{9}{\milli\tesla}$)\cite{Mitra_2017} on YAG substrates.  

The dynamic magnetic properties were investigated by broadband ferromagnetic resonance (bb-FMR) measurements.
\begin{figure}[htbp]%
	\includegraphics{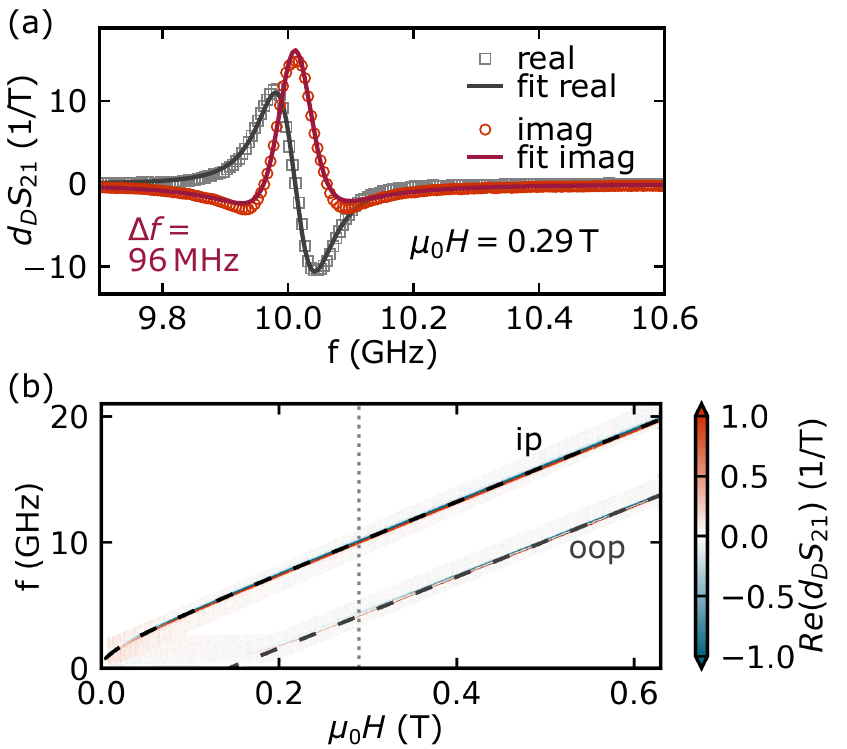}%
	\caption{The real and imaginary part of $d_{D}S_{21}$ for an external magnetic field of $\mu_0H=\SI{0.29}{\tesla}$ applied in the film plane are given in panel (a) as well as fits to the data following Ref.~[28]. False color plots of the real part of $d_{\mathrm{D}}S_{21}$ as a function of $\mu_0H$ for ip and oop alignment of the magnetic field as well as the Kittel fits\cite{Kittel_1948} to the respective data sets are depicted in panel (b).}%
	\label{ALD_YIG:fmr}%
\end{figure}%
For the bb-FMR measurements, the sample was mounted on a coplanar waveguide (CPW), which was connected to the two ports of a Vector Network Analyser N5225B (VNA) from Keysight. The CPW was then placed in the magnetic field generated by an electromagnet so that the magnetic field is either parallel or perpendicular to the substrate surface, which is denoted as in-plane (ip) and out-of-plane (oop) mounting, respectively. The complex microwave transmission parameter $S_{21}$ was measured with the VNA as a function of frequency $f$ (\SI{0.1}{\giga\hertz} - \SI{50}{\giga\hertz}) and the static magnetic field $\mu_0H$ (\SI{0}{\tesla} - \SI{1.9}{\tesla}) stepped with an increment of $\mu_0\Delta H=\SI{3.75}{\milli\tesla}$.

To remove the frequency dependent background signal of $S_{21}$ the derivative devide method was used.\cite{Maier_Flaig_2018} In Fig.~\ref{ALD_YIG:fmr}, panel(a) the the real and imaginary part of the magnetic field derivative of the complex microwave transmission parameter, i.e. $d_\mathrm{D} S_{21}$,\cite{Maier_Flaig_2018} are given for an in-plane (ip) external magnetic field of $\mu_0H=\SI{0.29}{\tesla}$. 
To determine the resonance frequency $f_{\mathrm{res}}$ and the FWHM linewidth $\Delta f$ at a fixed external magnetic field $H$, the complex $d_{\mathrm{D}}S_{21}$ data (cp. Fig.~\ref{ALD_YIG:fmr}, panel (a)) were fitted following Ref.~[28].
Please note, that the applied fitting procedure is only valid for neglecting magnetic anisotropies of higher order than uniaxial.
Fitting the data sets according to Ref.~[28] yields a FWHM linewidth of $\Delta f_{\mathrm{ip}}=\SI{96}{\mega\hertz}$ for $\Delta f_{\mathrm{res}}=\SI{10}{\giga\hertz}$. 

The full bb-FMR spectra for magnetic fields applied in-plane and along the surface normal (oop) are shown in Fig.~\ref{ALD_YIG:fmr}, panel (b). The g-factors as well as the effective magnetizations were extracted by fitting the data with the Kittel equations for thin films, which relate the resonance frequency $f_{\mathrm{res}}$ to the external magnetic field and the effective magnetization of the thin film.\cite{Kittel_1948} Note, that for the Kittel equations the magnetization and the external magnetic field are assumed to be pointing into the same direction and no crystal anisotropies are considered.
A g-factor of $g_{\mathrm{e,ip}}=2.02$ and an effective magnetization of $M_{\mathrm{eff,ip}}=\SI{115}{\kilo\ampere\per\meter}$ were extracted by the fit to the ip data. Fitting the oop data results in a g-factor of $g_{\mathrm{e,oop}}=2.00$ with an effective magnetization of $M_{\mathrm{eff, oop}}=\SI{114}{\kilo\ampere\per\meter}$, which is comparable to values reported in literature for other deposition techniques.\cite{Onbasli_2014} The g-factors are close to 2 as expected for YIG.
Using the g-factor determined from the Kittel fits, the frequency line width given in panel (b) can be converted into a magnetic field line width by $\Delta B=\frac{2\pi\hbar}{g_{\mathrm{e}}\mu_\mathrm{B}}\Delta f$. This results in $\Delta B_{\mathrm{ip}}=\SI{3.4}{\milli\tesla}$ at $f_{\mathrm{res}}=\SI{10}{\giga\hertz}$, which is comparable to values reported for thin films fabricated by pulsed laser deposition ($\Delta B=0.2-\SI{1.8}{\milli\tesla}$)\cite{Onbasli_2014, Sun_2012} and sputtering ($\Delta B=0.6-\SI{7.0}{\milli\tesla}$).\cite{Liu_2014,Yamamoto_2004}

\begin{figure}[htbp]%
	\includegraphics{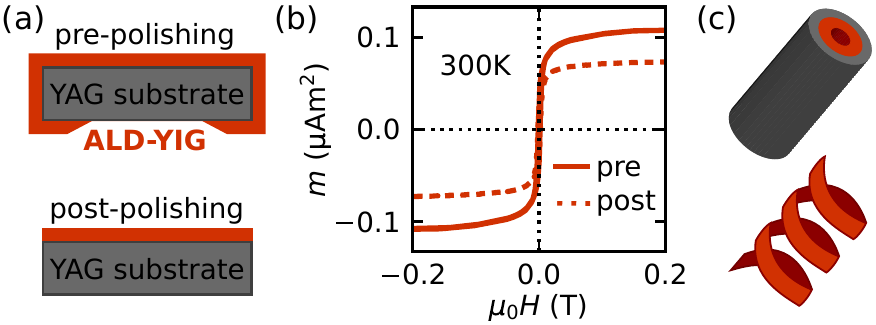}%
	\caption{A schematic of a cut through the pre- and post-polished sample is given in panel (a). The field dependent magnetic moment of the sample pre- and post polishing recorded at \SI{300}{\kelvin} with $H$ within the sample plane is shown in panel (b). In panel (c) possible geometries for future magnetic structures fabricated by ALD are given, where the ALD-YIG layer is indicated in orange.}%
	\label{ALD_YIG:squid_3D}%
\end{figure}

We finally turn to validating the conformal coating capability of the ALD process. Due to this generic feature of ALD, the nanolaminate is deposited not only on top of the substrates, but also on the sides, which is schematically depicted in Fig.~\ref{ALD_YIG:squid_3D}, panel~(a).
To a small extent, the ALD process might even result in a deposition on the backside of the substrate, which rests on the bottom of the deposition chamber. Thus, in such a conformally coated sample, a reduction of the magnetic moment upon polishing the side and back surfaces is expected. SQUID magnetometry data recorded pre- and post-polishing of the sample with $\ch{Y}:\ch{Fe}=0.6$ annealed at \SI{800}{\degreeCelsius} are given in panel (b) of Fig.~\ref{ALD_YIG:squid_3D}. Again, the linear background of the diamagnetic substrate is subtracted from the raw data. The magnetometry data verify that the measured moment is lowered by the polishing. This reduction can be quantified by calculating the change of the volume ratio $V_{\mathrm{pre}}/V_{\mathrm{post}}=1.41$ when considering a complete coating of the sides of the substrate, which agrees well with the measured change of the magnetic moment ratio of $m_{\mathrm{pre}}/m_{\mathrm{post}}=1.47$. The good agreement of these two ratios further corroborates that our ALD-YIG on the YAG substrate indeed forms a 3d magnetic structure. Using the post-polishing magnetometry data, we obtain a saturation magnetization of $M_{\mathrm{sat}}=\SI{133}{\kilo\ampere\per\meter}$, which is in good agreement with the saturation magnetization reported for thin films fabricated by pulsed laser deposition\cite{Althammer_2013} and sputter deposition with post annealing.\cite{Ma_2017,Hauser_2017} Please note, that the magnitude of the saturation magnetization also contradicts that the magnetic moment stems from \ch{Fe3O4}, $\gamma$-\ch{Fe2O3}, $\alpha$-\ch{Fe2O3} or \ch{YFeO3} as these are substantially different ($M_{\mathrm{sat},\mathrm{\ch{Fe3O4}}}=\SI{477}{\kilo\ampere\per\meter}$, $M_{\mathrm{sat},\gamma\text{-}\mathrm{\ch{Fe2O3}}}=\SI{430}{\kilo\ampere\per\meter}$, $M_{\mathrm{sat},\alpha\text{-}\mathrm{\ch{Fe2O3}}}=\SI{2}{\kilo\ampere\per\meter}$, $M_{\mathrm{sat},\ch{YFeO3}}=\SI{2}{\kilo\ampere\per\meter}$).\cite{Coey_2001,Shen_2009,Rosales_2019} 

By validating the conformity of our ALD-YIG films, we demonstrate the usability of the ALD-YIG thin films for 3D applications. For example coating a cylindrical YAG hole instead of a flat substrate would result in a YIG-cylinder, which is schematically shown in Fig.~\ref{ALD_YIG:squid_3D}, panel (c). Apart from that, also structures with a more complex topology such as spirals or even Möbius band-type structures are conceivable. In such structures new magnetic states are expected to manifest due to the non-trivial topology and their curvature.\cite{Streubel_2016,Fernandez_Pacheco_2017} 
However, to be able to exploit the full potential of the ALD-YIG process and to build 3D structures on the nanometer scale, first the fabrication of nanometer sized templates from YAG needs to be accomplished. This is non-trivial, as the 3D templates need to be crystalline and lithographic procedures like etching are complicated on oxidic garnets. Nevertheless, in these magnetic 3D nanostructures, interesting magnetization configurations have been envisioned, which are not only the basis for novel magnetization textures but are also predicted to result in a variety of other novel phenomena, such as non-reciprocity of spin waves\cite{Otalora_2016}, magnetochiral effects\cite{Dietrich_2008,Sloika_2014} or Cherenkov-like behavior of magnons.\cite{Yan_2013} The possibility of atomic layer deposited YIG paves the way for the experimental investigation of such phenomena. 

\section{summary}
We have established the atomic layer deposition of yttrium iron garnet (\ch{Y3Fe5O12}, YIG) thin films. The process is based on a supercycle approach, in which \ch{Y2O3} and \ch{Fe2O3} layers are sequentially stacked to achieve a so-called nanolaminate with an atomic ratio of $\ch{Y}:\ch{Fe}$ of $3:5$. A subsequent annealing step was used to achieve an interdiffusion of the nanolaminate and to promote the formation of high quality, crystalline yttrium iron garnet (YIG). As for conventional planar YIG films, annealing at temperatures of at least $\SI{700}{\degreeCelsius}$ was necessary to induce crystallization. The successful formation of YIG has been observed for variations in the stoichiometry of up to $\pm\SI{30}{\percent}$, demonstrating a high stability of the nanolaminate approach with respect to process related deviations from the ideal chemical composition. Changing the number of monolayers in the supercycle process also did not hamper the successful crystallization into YIG. TEM studies revealed a high crystalline quality of the YIG layers after the annealing step.
The coercive field of $\mu_0H_{\mathrm{c}}=\SI{1.5}{\milli\tesla}$ as well as the saturation magnetization of $M_{\mathrm{sat}}=\SI{133}{\kilo\ampere\per\meter}$ at \SI{300}{\kelvin} are in good agreement with values reported for other deposition techniques.\cite{Althammer_2013,Ma_2017} Broadband ferromagnetic resonance experiments yield a g-factor of $g=2$ and an effective magnetization of $M_{\mathrm{eff}}=\SI{115}{\kilo\ampere\per\meter}$, in good agreement with the literature.\cite{Onbasli_2014} Also the FMR linewidth in the magnetic field domain $\Delta B_{\mathrm{ip}}=\SI{3.4}{\milli\tesla}$ at \SI{10}{\giga\hertz} is comparable to the one of thin films fabricated by other deposition techniques.\cite{Onbasli_2014,Liu_2014}	
Importantly, we also validated the conformity of the ALD process for YIG by showing that not only the surface, but also the sides of the substrate indeed were coated in the process. Therewith, our results open the door to using one of the most interesting magnetic insulators for spintronics in future experiments on three dimensional (nano-)structures.

\section{Acknowledgements}
The authors thank Almut Pöhl from the IFW Dresden for the sample preparation for the transmission electron microscopy measurements.

\bibliography{references_ALD_YIG.bib}

\begin{thebibliography}{44}%
\makeatletter
\providecommand \@ifxundefined [1]{%
 \@ifx{#1\undefined}
}%
\providecommand \@ifnum [1]{%
 \ifnum #1\expandafter \@firstoftwo
 \else \expandafter \@secondoftwo
 \fi
}%
\providecommand \@ifx [1]{%
 \ifx #1\expandafter \@firstoftwo
 \else \expandafter \@secondoftwo
 \fi
}%
\providecommand \natexlab [1]{#1}%
\providecommand \enquote  [1]{``#1''}%
\providecommand \bibnamefont  [1]{#1}%
\providecommand \bibfnamefont [1]{#1}%
\providecommand \citenamefont [1]{#1}%
\providecommand \href@noop [0]{\@secondoftwo}%
\providecommand \href [0]{\begingroup \@sanitize@url \@href}%
\providecommand \@href[1]{\@@startlink{#1}\@@href}%
\providecommand \@@href[1]{\endgroup#1\@@endlink}%
\providecommand \@sanitize@url [0]{\catcode `\\12\catcode `\$12\catcode
  `\&12\catcode `\#12\catcode `\^12\catcode `\_12\catcode `\%12\relax}%
\providecommand \@@startlink[1]{}%
\providecommand \@@endlink[0]{}%
\providecommand \url  [0]{\begingroup\@sanitize@url \@url }%
\providecommand \@url [1]{\endgroup\@href {#1}{\urlprefix }}%
\providecommand \urlprefix  [0]{URL }%
\providecommand \Eprint [0]{\href }%
\providecommand \doibase [0]{https://doi.org/}%
\providecommand \selectlanguage [0]{\@gobble}%
\providecommand \bibinfo  [0]{\@secondoftwo}%
\providecommand \bibfield  [0]{\@secondoftwo}%
\providecommand \translation [1]{[#1]}%
\providecommand \BibitemOpen [0]{}%
\providecommand \bibitemStop [0]{}%
\providecommand \bibitemNoStop [0]{.\EOS\space}%
\providecommand \EOS [0]{\spacefactor3000\relax}%
\providecommand \BibitemShut  [1]{\csname bibitem#1\endcsname}%
\let\auto@bib@innerbib\@empty
%</preamble>
\bibitem [{\citenamefont {Parkin}\ \emph {et~al.}(2008)\citenamefont {Parkin},
  \citenamefont {Hayashi},\ and\ \citenamefont {Thomas}}]{Parkin_2008}%
  \BibitemOpen
  \bibfield  {author} {\bibinfo {author} {\bibfnamefont {S.~S.~P.}\
  \bibnamefont {Parkin}}, \bibinfo {author} {\bibfnamefont {M.}~\bibnamefont
  {Hayashi}},\ and\ \bibinfo {author} {\bibfnamefont {L.}~\bibnamefont
  {Thomas}},\ }\bibfield  {title} {\bibinfo {title} {Magnetic domain-wall
  racetrack memory},\ }\href {https://doi.org/10.1126/science.1145799}
  {\bibfield  {journal} {\bibinfo  {journal} {Science}\ }\textbf {\bibinfo
  {volume} {320}},\ \bibinfo {pages} {190} (\bibinfo {year}
  {2008})}\BibitemShut {NoStop}%
\bibitem [{\citenamefont {Streubel}\ \emph {et~al.}(2016)\citenamefont
  {Streubel}, \citenamefont {Fischer}, \citenamefont {Kronast}, \citenamefont
  {Kravchuk}, \citenamefont {Sheka}, \citenamefont {Gaididei}, \citenamefont
  {Schmidt},\ and\ \citenamefont {Makarov}}]{Streubel_2016}%
  \BibitemOpen
  \bibfield  {author} {\bibinfo {author} {\bibfnamefont {R.}~\bibnamefont
  {Streubel}}, \bibinfo {author} {\bibfnamefont {P.}~\bibnamefont {Fischer}},
  \bibinfo {author} {\bibfnamefont {F.}~\bibnamefont {Kronast}}, \bibinfo
  {author} {\bibfnamefont {V.~P.}\ \bibnamefont {Kravchuk}}, \bibinfo {author}
  {\bibfnamefont {D.~D.}\ \bibnamefont {Sheka}}, \bibinfo {author}
  {\bibfnamefont {Y.}~\bibnamefont {Gaididei}}, \bibinfo {author}
  {\bibfnamefont {O.~G.}\ \bibnamefont {Schmidt}},\ and\ \bibinfo {author}
  {\bibfnamefont {D.}~\bibnamefont {Makarov}},\ }\bibfield  {title} {\bibinfo
  {title} {Magnetism in curved geometries},\ }\href
  {https://doi.org/10.1088/0022-3727/49/36/363001} {\bibfield  {journal}
  {\bibinfo  {journal} {Journal of Physics D: Applied Physics}\ }\textbf
  {\bibinfo {volume} {49}},\ \bibinfo {pages} {363001} (\bibinfo {year}
  {2016})}\BibitemShut {NoStop}%
\bibitem [{\citenamefont {Fern{\'{a}}ndez-Pacheco}\ \emph
  {et~al.}(2017)\citenamefont {Fern{\'{a}}ndez-Pacheco}, \citenamefont
  {Streubel}, \citenamefont {Fruchart}, \citenamefont {Hertel}, \citenamefont
  {Fischer},\ and\ \citenamefont {Cowburn}}]{Fernandez_Pacheco_2017}%
  \BibitemOpen
  \bibfield  {author} {\bibinfo {author} {\bibfnamefont {A.}~\bibnamefont
  {Fern{\'{a}}ndez-Pacheco}}, \bibinfo {author} {\bibfnamefont
  {R.}~\bibnamefont {Streubel}}, \bibinfo {author} {\bibfnamefont
  {O.}~\bibnamefont {Fruchart}}, \bibinfo {author} {\bibfnamefont
  {R.}~\bibnamefont {Hertel}}, \bibinfo {author} {\bibfnamefont
  {P.}~\bibnamefont {Fischer}},\ and\ \bibinfo {author} {\bibfnamefont {R.~P.}\
  \bibnamefont {Cowburn}},\ }\bibfield  {title} {\bibinfo {title}
  {Three-dimensional nanomagnetism},\ }\href@noop {} {\bibfield  {journal}
  {\bibinfo  {journal} {Nature Communications}\ }\textbf {\bibinfo {volume}
  {8}} (\bibinfo {year} {2017})}\BibitemShut {NoStop}%
\bibitem [{\citenamefont {Vedmedenko}\ \emph {et~al.}(2020)\citenamefont
  {Vedmedenko}, \citenamefont {Kawakami}, \citenamefont {Sheka}, \citenamefont
  {Gambardella}, \citenamefont {Kirilyuk}, \citenamefont {Hirohata},
  \citenamefont {Binek}, \citenamefont {Chubykalo-Fesenko}, \citenamefont
  {Sanvito}, \citenamefont {Kirby}, \citenamefont {Grollier}, \citenamefont
  {Everschor-Sitte}, \citenamefont {Kampfrath}, \citenamefont {You},\ and\
  \citenamefont {Berger}}]{Vedmedenko_2020}%
  \BibitemOpen
  \bibfield  {author} {\bibinfo {author} {\bibfnamefont {E.~Y.}\ \bibnamefont
  {Vedmedenko}}, \bibinfo {author} {\bibfnamefont {R.~K.}\ \bibnamefont
  {Kawakami}}, \bibinfo {author} {\bibfnamefont {D.~D.}\ \bibnamefont {Sheka}},
  \bibinfo {author} {\bibfnamefont {P.}~\bibnamefont {Gambardella}}, \bibinfo
  {author} {\bibfnamefont {A.}~\bibnamefont {Kirilyuk}}, \bibinfo {author}
  {\bibfnamefont {A.}~\bibnamefont {Hirohata}}, \bibinfo {author}
  {\bibfnamefont {C.}~\bibnamefont {Binek}}, \bibinfo {author} {\bibfnamefont
  {O.}~\bibnamefont {Chubykalo-Fesenko}}, \bibinfo {author} {\bibfnamefont
  {S.}~\bibnamefont {Sanvito}}, \bibinfo {author} {\bibfnamefont {B.~J.}\
  \bibnamefont {Kirby}}, \bibinfo {author} {\bibfnamefont {J.}~\bibnamefont
  {Grollier}}, \bibinfo {author} {\bibfnamefont {K.}~\bibnamefont
  {Everschor-Sitte}}, \bibinfo {author} {\bibfnamefont {T.}~\bibnamefont
  {Kampfrath}}, \bibinfo {author} {\bibfnamefont {C.-Y.}\ \bibnamefont {You}},\
  and\ \bibinfo {author} {\bibfnamefont {A.}~\bibnamefont {Berger}},\
  }\bibfield  {title} {\bibinfo {title} {The 2020 magnetism roadmap},\ }\href
  {https://doi.org/10.1088/1361-6463/ab9d98} {\bibfield  {journal} {\bibinfo
  {journal} {Journal of Physics D: Applied Physics}\ }\textbf {\bibinfo
  {volume} {53}},\ \bibinfo {pages} {453001} (\bibinfo {year}
  {2020})}\BibitemShut {NoStop}%
\bibitem [{\citenamefont {Das}\ \emph {et~al.}(2019)\citenamefont {Das},
  \citenamefont {Makarov}, \citenamefont {Gentile}, \citenamefont {Cuoco},
  \citenamefont {van Wees}, \citenamefont {Ortix},\ and\ \citenamefont
  {Vera-Marun}}]{Das_2019}%
  \BibitemOpen
  \bibfield  {author} {\bibinfo {author} {\bibfnamefont {K.~S.}\ \bibnamefont
  {Das}}, \bibinfo {author} {\bibfnamefont {D.}~\bibnamefont {Makarov}},
  \bibinfo {author} {\bibfnamefont {P.}~\bibnamefont {Gentile}}, \bibinfo
  {author} {\bibfnamefont {M.}~\bibnamefont {Cuoco}}, \bibinfo {author}
  {\bibfnamefont {B.~J.}\ \bibnamefont {van Wees}}, \bibinfo {author}
  {\bibfnamefont {C.}~\bibnamefont {Ortix}},\ and\ \bibinfo {author}
  {\bibfnamefont {I.~J.}\ \bibnamefont {Vera-Marun}},\ }\bibfield  {title}
  {\bibinfo {title} {Independent geometrical control of spin and charge
  resistances in curved spintronics},\ }\href
  {https://doi.org/10.1021/acs.nanolett.9b01994} {\bibfield  {journal}
  {\bibinfo  {journal} {Nano Letters}\ }\textbf {\bibinfo {volume} {19}},\
  \bibinfo {pages} {6839} (\bibinfo {year} {2019})}\BibitemShut {NoStop}%
\bibitem [{\citenamefont {Sheka}\ \emph {et~al.}(2020)\citenamefont {Sheka},
  \citenamefont {Pylypovskyi}, \citenamefont {Landeros}, \citenamefont
  {Gaididei}, \citenamefont {K{\'{a}}kay},\ and\ \citenamefont
  {Makarov}}]{Sheka_2020}%
  \BibitemOpen
  \bibfield  {author} {\bibinfo {author} {\bibfnamefont {D.~D.}\ \bibnamefont
  {Sheka}}, \bibinfo {author} {\bibfnamefont {O.~V.}\ \bibnamefont
  {Pylypovskyi}}, \bibinfo {author} {\bibfnamefont {P.}~\bibnamefont
  {Landeros}}, \bibinfo {author} {\bibfnamefont {Y.}~\bibnamefont {Gaididei}},
  \bibinfo {author} {\bibfnamefont {A.}~\bibnamefont {K{\'{a}}kay}},\ and\
  \bibinfo {author} {\bibfnamefont {D.}~\bibnamefont {Makarov}},\ }\bibfield
  {title} {\bibinfo {title} {Nonlocal chiral symmetry breaking in curvilinear
  magnetic shells},\ }\href@noop {} {\bibfield  {journal} {\bibinfo  {journal}
  {Communications Physics}\ }\textbf {\bibinfo {volume} {3}} (\bibinfo {year}
  {2020})}\BibitemShut {NoStop}%
\bibitem [{\citenamefont {Ot{\'{a}}lora}\ \emph {et~al.}(2016)\citenamefont
  {Ot{\'{a}}lora}, \citenamefont {Yan}, \citenamefont {Schultheiss},
  \citenamefont {Hertel},\ and\ \citenamefont {K{\'{a}}kay}}]{Otalora_2016}%
  \BibitemOpen
  \bibfield  {author} {\bibinfo {author} {\bibfnamefont {J.~A.}\ \bibnamefont
  {Ot{\'{a}}lora}}, \bibinfo {author} {\bibfnamefont {M.}~\bibnamefont {Yan}},
  \bibinfo {author} {\bibfnamefont {H.}~\bibnamefont {Schultheiss}}, \bibinfo
  {author} {\bibfnamefont {R.}~\bibnamefont {Hertel}},\ and\ \bibinfo {author}
  {\bibfnamefont {A.}~\bibnamefont {K{\'{a}}kay}},\ }\bibfield  {title}
  {\bibinfo {title} {Curvature-induced asymmetric spin-wave dispersion},\
  }\href@noop {} {\bibfield  {journal} {\bibinfo  {journal} {Physical Review
  Letters}\ }\textbf {\bibinfo {volume} {117}} (\bibinfo {year}
  {2016})}\BibitemShut {NoStop}%
\bibitem [{\citenamefont {Yan}\ \emph {et~al.}(2013)\citenamefont {Yan},
  \citenamefont {K{\'{a}}kay}, \citenamefont {Andreas},\ and\ \citenamefont
  {Hertel}}]{Yan_2013}%
  \BibitemOpen
  \bibfield  {author} {\bibinfo {author} {\bibfnamefont {M.}~\bibnamefont
  {Yan}}, \bibinfo {author} {\bibfnamefont {A.}~\bibnamefont {K{\'{a}}kay}},
  \bibinfo {author} {\bibfnamefont {C.}~\bibnamefont {Andreas}},\ and\ \bibinfo
  {author} {\bibfnamefont {R.}~\bibnamefont {Hertel}},\ }\bibfield  {title}
  {\bibinfo {title} {Spin-cherenkov effect and magnonic mach cones},\
  }\href@noop {} {\bibfield  {journal} {\bibinfo  {journal} {Physical Review
  B}\ }\textbf {\bibinfo {volume} {88}} (\bibinfo {year} {2013})}\BibitemShut
  {NoStop}%
\bibitem [{\citenamefont {Miikkulainen}\ \emph {et~al.}(2013)\citenamefont
  {Miikkulainen}, \citenamefont {Leskelä}, \citenamefont {Ritala},\ and\
  \citenamefont {Puurunen}}]{Miikkulainen_2013}%
  \BibitemOpen
  \bibfield  {author} {\bibinfo {author} {\bibfnamefont {V.}~\bibnamefont
  {Miikkulainen}}, \bibinfo {author} {\bibfnamefont {M.}~\bibnamefont
  {Leskelä}}, \bibinfo {author} {\bibfnamefont {M.}~\bibnamefont {Ritala}},\
  and\ \bibinfo {author} {\bibfnamefont {R.~L.}\ \bibnamefont {Puurunen}},\
  }\bibfield  {title} {\bibinfo {title} {Crystallinity of inorganic films grown
  by atomic layer deposition: Overview and general trends},\ }\href@noop {}
  {\bibfield  {journal} {\bibinfo  {journal} {Journal of Applied Physics}\
  }\textbf {\bibinfo {volume} {113}},\ \bibinfo {pages} {021301} (\bibinfo
  {year} {2013})}\BibitemShut {NoStop}%
\bibitem [{\citenamefont {Elers}\ \emph {et~al.}(2006)\citenamefont {Elers},
  \citenamefont {Blomberg}, \citenamefont {Peussa}, \citenamefont {Aitchison},
  \citenamefont {Haukka},\ and\ \citenamefont {Marcus}}]{Elers_2006}%
  \BibitemOpen
  \bibfield  {author} {\bibinfo {author} {\bibfnamefont {K.-E.}\ \bibnamefont
  {Elers}}, \bibinfo {author} {\bibfnamefont {T.}~\bibnamefont {Blomberg}},
  \bibinfo {author} {\bibfnamefont {M.}~\bibnamefont {Peussa}}, \bibinfo
  {author} {\bibfnamefont {B.}~\bibnamefont {Aitchison}}, \bibinfo {author}
  {\bibfnamefont {S.}~\bibnamefont {Haukka}},\ and\ \bibinfo {author}
  {\bibfnamefont {S.}~\bibnamefont {Marcus}},\ }\bibfield  {title} {\bibinfo
  {title} {Film uniformity in atomic layer deposition},\ }\href
  {https://doi.org/10.1002/cvde.200500024} {\bibfield  {journal} {\bibinfo
  {journal} {Chemical Vapor Deposition}\ }\textbf {\bibinfo {volume} {12}},\
  \bibinfo {pages} {13} (\bibinfo {year} {2006})}\BibitemShut {NoStop}%
\bibitem [{\citenamefont {Mackus}\ \emph {et~al.}(2018)\citenamefont {Mackus},
  \citenamefont {Schneider}, \citenamefont {MacIsaac}, \citenamefont {Baker},\
  and\ \citenamefont {Bent}}]{Mackus_2018}%
  \BibitemOpen
  \bibfield  {author} {\bibinfo {author} {\bibfnamefont {A.~J.~M.}\
  \bibnamefont {Mackus}}, \bibinfo {author} {\bibfnamefont {J.~R.}\
  \bibnamefont {Schneider}}, \bibinfo {author} {\bibfnamefont {C.}~\bibnamefont
  {MacIsaac}}, \bibinfo {author} {\bibfnamefont {J.~G.}\ \bibnamefont
  {Baker}},\ and\ \bibinfo {author} {\bibfnamefont {S.~F.}\ \bibnamefont
  {Bent}},\ }\bibfield  {title} {\bibinfo {title} {Synthesis of doped, ternary,
  and quaternary materials by atomic layer deposition: A review},\ }\href
  {https://doi.org/10.1021/acs.chemmater.8b02878} {\bibfield  {journal}
  {\bibinfo  {journal} {Chemistry of Materials}\ }\textbf {\bibinfo {volume}
  {31}},\ \bibinfo {pages} {1142} (\bibinfo {year} {2018})}\BibitemShut
  {NoStop}%
\bibitem [{\citenamefont {Althammer}\ \emph {et~al.}(2013)\citenamefont
  {Althammer}, \citenamefont {Meyer}, \citenamefont {Nakayama}, \citenamefont
  {Schreier}, \citenamefont {Altmannshofer}, \citenamefont {Weiler},
  \citenamefont {Huebl}, \citenamefont {Gepr\"ags}, \citenamefont {Opel},
  \citenamefont {Gross}, \citenamefont {Meier}, \citenamefont {Klewe},
  \citenamefont {Kuschel}, \citenamefont {Schmalhorst}, \citenamefont {Reiss},
  \citenamefont {Shen}, \citenamefont {Gupta}, \citenamefont {Chen},
  \citenamefont {Bauer}, \citenamefont {Saitoh},\ and\ \citenamefont
  {Goennenwein}}]{Althammer_2013}%
  \BibitemOpen
  \bibfield  {author} {\bibinfo {author} {\bibfnamefont {M.}~\bibnamefont
  {Althammer}}, \bibinfo {author} {\bibfnamefont {S.}~\bibnamefont {Meyer}},
  \bibinfo {author} {\bibfnamefont {H.}~\bibnamefont {Nakayama}}, \bibinfo
  {author} {\bibfnamefont {M.}~\bibnamefont {Schreier}}, \bibinfo {author}
  {\bibfnamefont {S.}~\bibnamefont {Altmannshofer}}, \bibinfo {author}
  {\bibfnamefont {M.}~\bibnamefont {Weiler}}, \bibinfo {author} {\bibfnamefont
  {H.}~\bibnamefont {Huebl}}, \bibinfo {author} {\bibfnamefont
  {S.}~\bibnamefont {Gepr\"ags}}, \bibinfo {author} {\bibfnamefont
  {M.}~\bibnamefont {Opel}}, \bibinfo {author} {\bibfnamefont {R.}~\bibnamefont
  {Gross}}, \bibinfo {author} {\bibfnamefont {D.}~\bibnamefont {Meier}},
  \bibinfo {author} {\bibfnamefont {C.}~\bibnamefont {Klewe}}, \bibinfo
  {author} {\bibfnamefont {T.}~\bibnamefont {Kuschel}}, \bibinfo {author}
  {\bibfnamefont {J.-M.}\ \bibnamefont {Schmalhorst}}, \bibinfo {author}
  {\bibfnamefont {G.}~\bibnamefont {Reiss}}, \bibinfo {author} {\bibfnamefont
  {L.}~\bibnamefont {Shen}}, \bibinfo {author} {\bibfnamefont {A.}~\bibnamefont
  {Gupta}}, \bibinfo {author} {\bibfnamefont {Y.-T.}\ \bibnamefont {Chen}},
  \bibinfo {author} {\bibfnamefont {G.~E.~W.}\ \bibnamefont {Bauer}}, \bibinfo
  {author} {\bibfnamefont {E.}~\bibnamefont {Saitoh}},\ and\ \bibinfo {author}
  {\bibfnamefont {S.~T.~B.}\ \bibnamefont {Goennenwein}},\ }\bibfield  {title}
  {\bibinfo {title} {Quantitative study of the spin hall magnetoresistance in
  ferromagnetic insulator/normal metal hybrids},\ }\href
  {https://doi.org/10.1103/PhysRevB.87.224401} {\bibfield  {journal} {\bibinfo
  {journal} {Phys. Rev. B}\ }\textbf {\bibinfo {volume} {87}},\ \bibinfo
  {pages} {224401} (\bibinfo {year} {2013})}\BibitemShut {NoStop}%
\bibitem [{\citenamefont {Hauser}\ \emph {et~al.}(2017)\citenamefont {Hauser},
  \citenamefont {Eisenschmidt}, \citenamefont {Richter}, \citenamefont
  {Müller}, \citenamefont {Deniz},\ and\ \citenamefont
  {Schmidt}}]{Hauser_2017}%
  \BibitemOpen
  \bibfield  {author} {\bibinfo {author} {\bibfnamefont {C.}~\bibnamefont
  {Hauser}}, \bibinfo {author} {\bibfnamefont {C.}~\bibnamefont
  {Eisenschmidt}}, \bibinfo {author} {\bibfnamefont {T.}~\bibnamefont
  {Richter}}, \bibinfo {author} {\bibfnamefont {A.}~\bibnamefont {Müller}},
  \bibinfo {author} {\bibfnamefont {H.}~\bibnamefont {Deniz}},\ and\ \bibinfo
  {author} {\bibfnamefont {G.}~\bibnamefont {Schmidt}},\ }\bibfield  {title}
  {\bibinfo {title} {Annealing of amorphous yttrium iron garnet thin films in
  argon atmosphere},\ }\href {https://doi.org/10.1063/1.4999829} {\bibfield
  {journal} {\bibinfo  {journal} {Journal of Applied Physics}\ }\textbf
  {\bibinfo {volume} {122}},\ \bibinfo {pages} {083908} (\bibinfo {year}
  {2017})}\BibitemShut {NoStop}%
\bibitem [{\citenamefont {Askeland}(1996)}]{Askeland_1996}%
  \BibitemOpen
  \bibfield  {author} {\bibinfo {author} {\bibfnamefont {D.~R.}\ \bibnamefont
  {Askeland}},\ }\href {https://doi.org/10.1007/978-1-4899-2895-5} {\emph
  {\bibinfo {title} {The Science and Engineering of Materials}}}\ (\bibinfo
  {publisher} {Springer {US}},\ \bibinfo {year} {1996})\BibitemShut {NoStop}%
\bibitem [{\citenamefont {Villars}\ and\ \citenamefont
  {Cenzual}(2016)}]{Villars_2016_Y2O3}%
  \BibitemOpen
  \bibinfo {editor} {\bibfnamefont {P.}~\bibnamefont {Villars}}\ and\ \bibinfo
  {editor} {\bibfnamefont {K.}~\bibnamefont {Cenzual}},\ eds.,\ \href
  {https://materials.springer.com/isp/crystallographic/docs/sd{\_}1923708}
  {\emph {\bibinfo {title} {Y$_2$O$_3$ Crystal Structure: Datasheet from
  "PAULING FILE Multinaries Edition - 2012" in SpringerMaterials}}}\ (\bibinfo
  {publisher} {Springer-Verlag Berlin Heidelberg},\ \bibinfo {year}
  {2016})\BibitemShut {NoStop}%
\bibitem [{\citenamefont {Madelung}\ \emph {et~al.}(2000)\citenamefont
  {Madelung}, \citenamefont {R\"ossler},\ and\ \citenamefont
  {Schulz}}]{Landolt-Bornstein:hematite_2000}%
  \BibitemOpen
  \bibinfo {editor} {\bibfnamefont {O.}~\bibnamefont {Madelung}}, \bibinfo
  {editor} {\bibfnamefont {U.}~\bibnamefont {R\"ossler}},\ and\ \bibinfo
  {editor} {\bibfnamefont {M.}~\bibnamefont {Schulz}},\ eds.,\ \href
  {https://doi.org/10.1007/10681735_543} {\emph {\bibinfo {title} {Hematite
  (alpha-Fe$_2$O$_3$), {D}atasheet from {L}andolt-{B}\"ornstein}}}\ (\bibinfo
  {publisher} {Springer-Verlag Berlin Heidelberg},\ \bibinfo {year}
  {2000})\BibitemShut {NoStop}%
\bibitem [{\citenamefont {Clementi}\ and\ \citenamefont
  {Raimondi}(1963)}]{Clementi_1963}%
  \BibitemOpen
  \bibfield  {author} {\bibinfo {author} {\bibfnamefont {E.}~\bibnamefont
  {Clementi}}\ and\ \bibinfo {author} {\bibfnamefont {D.~L.}\ \bibnamefont
  {Raimondi}},\ }\bibfield  {title} {\bibinfo {title} {Atomic screening
  constants from scf functions},\ }\href {https://doi.org/10.1063/1.1733573}
  {\bibfield  {journal} {\bibinfo  {journal} {The Journal of Chemical Physics}\
  }\textbf {\bibinfo {volume} {38}},\ \bibinfo {pages} {2686} (\bibinfo {year}
  {1963})}\BibitemShut {NoStop}%
\bibitem [{\citenamefont {Slater}(1964)}]{Slater_1964}%
  \BibitemOpen
  \bibfield  {author} {\bibinfo {author} {\bibfnamefont {J.~C.}\ \bibnamefont
  {Slater}},\ }\bibfield  {title} {\bibinfo {title} {Atomic radii in
  crystals},\ }\href {https://doi.org/10.1063/1.1725697} {\bibfield  {journal}
  {\bibinfo  {journal} {The Journal of Chemical Physics}\ }\textbf {\bibinfo
  {volume} {41}},\ \bibinfo {pages} {3199} (\bibinfo {year}
  {1964})}\BibitemShut {NoStop}%
\bibitem [{\citenamefont {Wiegand}\ \emph {et~al.}(2018)\citenamefont
  {Wiegand}, \citenamefont {Faust}, \citenamefont {Meinhardt}, \citenamefont
  {Blick}, \citenamefont {Zierold},\ and\ \citenamefont
  {Nielsch}}]{Wiegand_2018}%
  \BibitemOpen
  \bibfield  {author} {\bibinfo {author} {\bibfnamefont {C.~W.}\ \bibnamefont
  {Wiegand}}, \bibinfo {author} {\bibfnamefont {R.}~\bibnamefont {Faust}},
  \bibinfo {author} {\bibfnamefont {A.}~\bibnamefont {Meinhardt}}, \bibinfo
  {author} {\bibfnamefont {R.~H.}\ \bibnamefont {Blick}}, \bibinfo {author}
  {\bibfnamefont {R.}~\bibnamefont {Zierold}},\ and\ \bibinfo {author}
  {\bibfnamefont {K.}~\bibnamefont {Nielsch}},\ }\bibfield  {title} {\bibinfo
  {title} {Understanding the growth mechanisms of multilayered systems in
  atomic layer deposition process},\ }\href
  {https://doi.org/10.1021/acs.chemmater.7b05128} {\bibfield  {journal}
  {\bibinfo  {journal} {Chemistry of Materials}\ }\textbf {\bibinfo {volume}
  {30}},\ \bibinfo {pages} {1971} (\bibinfo {year} {2018})}\BibitemShut
  {NoStop}%
\bibitem [{\citenamefont {Myers}\ \emph {et~al.}(2021)\citenamefont {Myers},
  \citenamefont {Cano}, \citenamefont {Lancaster}, \citenamefont {Clancey},\
  and\ \citenamefont {George}}]{Myers_2021}%
  \BibitemOpen
  \bibfield  {author} {\bibinfo {author} {\bibfnamefont {T.~J.}\ \bibnamefont
  {Myers}}, \bibinfo {author} {\bibfnamefont {A.~M.}\ \bibnamefont {Cano}},
  \bibinfo {author} {\bibfnamefont {D.~K.}\ \bibnamefont {Lancaster}}, \bibinfo
  {author} {\bibfnamefont {J.~W.}\ \bibnamefont {Clancey}},\ and\ \bibinfo
  {author} {\bibfnamefont {S.~M.}\ \bibnamefont {George}},\ }\bibfield  {title}
  {\bibinfo {title} {Conversion reactions in atomic layer processing with
  emphasis on {Z}n{O} conversion to {A}l$_2${O}$_3$ by trimethylaluminum},\
  }\href {https://doi.org/10.1116/6.0000680} {\bibfield  {journal} {\bibinfo
  {journal} {Journal of Vacuum Science \& Technology A}\ }\textbf {\bibinfo
  {volume} {39}},\ \bibinfo {pages} {021001} (\bibinfo {year}
  {2021})}\BibitemShut {NoStop}%
\bibitem [{\citenamefont {Lammel}\ \emph {et~al.}(2020)\citenamefont {Lammel},
  \citenamefont {Geishendorf}, \citenamefont {Choffel}, \citenamefont {Hamann},
  \citenamefont {Johnson}, \citenamefont {Nielsch},\ and\ \citenamefont
  {Thomas}}]{Lammel_2020}%
  \BibitemOpen
  \bibfield  {author} {\bibinfo {author} {\bibfnamefont {M.}~\bibnamefont
  {Lammel}}, \bibinfo {author} {\bibfnamefont {K.}~\bibnamefont {Geishendorf}},
  \bibinfo {author} {\bibfnamefont {M.~A.}\ \bibnamefont {Choffel}}, \bibinfo
  {author} {\bibfnamefont {D.~M.}\ \bibnamefont {Hamann}}, \bibinfo {author}
  {\bibfnamefont {D.~C.}\ \bibnamefont {Johnson}}, \bibinfo {author}
  {\bibfnamefont {K.}~\bibnamefont {Nielsch}},\ and\ \bibinfo {author}
  {\bibfnamefont {A.}~\bibnamefont {Thomas}},\ }\bibfield  {title} {\bibinfo
  {title} {Fast fourier transform and multi-gaussian fitting of {XRR} data to
  determine the thickness of {ALD} grown thin films within the initial growth
  regime},\ }\href {https://doi.org/10.1063/5.0024991} {\bibfield  {journal}
  {\bibinfo  {journal} {Applied Physics Letters}\ }\textbf {\bibinfo {volume}
  {117}},\ \bibinfo {pages} {213106} (\bibinfo {year} {2020})}\BibitemShut
  {NoStop}%
\bibitem [{\citenamefont {Lammel}(2021)}]{Lammel_PhD}%
  \BibitemOpen
  \bibfield  {author} {\bibinfo {author} {\bibfnamefont {M.}~\bibnamefont
  {Lammel}},\ }\emph {\bibinfo {title} {Advancing 3D Spintronics: Atomic Layer
  Deposition of Platinum and Yttrium Iron Garnet Thin Films}},\ \href@noop {}
  {Ph.D. thesis},\ \bibinfo  {school} {Technische Universität Dresden}
  (\bibinfo {year} {2021})\BibitemShut {NoStop}%
\bibitem [{\citenamefont {Kumar}\ \emph {et~al.}(2008)\citenamefont {Kumar},
  \citenamefont {Prasad}, \citenamefont {Misra}, \citenamefont {Venkataramani},
  \citenamefont {Bohra},\ and\ \citenamefont {Krishnan}}]{Kumar_2008}%
  \BibitemOpen
  \bibfield  {author} {\bibinfo {author} {\bibfnamefont {N.}~\bibnamefont
  {Kumar}}, \bibinfo {author} {\bibfnamefont {S.}~\bibnamefont {Prasad}},
  \bibinfo {author} {\bibfnamefont {D.}~\bibnamefont {Misra}}, \bibinfo
  {author} {\bibfnamefont {N.}~\bibnamefont {Venkataramani}}, \bibinfo {author}
  {\bibfnamefont {M.}~\bibnamefont {Bohra}},\ and\ \bibinfo {author}
  {\bibfnamefont {R.}~\bibnamefont {Krishnan}},\ }\bibfield  {title} {\bibinfo
  {title} {The influence of substrate temperature and annealing on the
  properties of pulsed laser-deposited {YIG} films on fused quartz substrate},\
  }\href {https://doi.org/10.1016/j.jmmm.2008.04.112} {\bibfield  {journal}
  {\bibinfo  {journal} {Journal of Magnetism and Magnetic Materials}\ }\textbf
  {\bibinfo {volume} {320}},\ \bibinfo {pages} {2233} (\bibinfo {year}
  {2008})}\BibitemShut {NoStop}%
\bibitem [{\citenamefont {Park}\ and\ \citenamefont {Cho}(2001)}]{Park_2001}%
  \BibitemOpen
  \bibfield  {author} {\bibinfo {author} {\bibfnamefont {M.-B.}\ \bibnamefont
  {Park}}\ and\ \bibinfo {author} {\bibfnamefont {N.-H.}\ \bibnamefont {Cho}},\
  }\bibfield  {title} {\bibinfo {title} {Structural and magnetic
  characteristics of yttrium iron garnet ({YIG}, ce:{YIG}) films prepared by
  {RF} magnetron sputter techniques},\ }\href
  {https://doi.org/10.1016/s0304-8853(01)00068-3} {\bibfield  {journal}
  {\bibinfo  {journal} {Journal of Magnetism and Magnetic Materials}\ }\textbf
  {\bibinfo {volume} {231}},\ \bibinfo {pages} {253} (\bibinfo {year}
  {2001})}\BibitemShut {NoStop}%
\bibitem [{\citenamefont {Yamamoto}\ \emph {et~al.}(2004)\citenamefont
  {Yamamoto}, \citenamefont {Kuniki}, \citenamefont {Kurisu}, \citenamefont
  {Matsuura},\ and\ \citenamefont {Jang}}]{Yamamoto_2004}%
  \BibitemOpen
  \bibfield  {author} {\bibinfo {author} {\bibfnamefont {S.}~\bibnamefont
  {Yamamoto}}, \bibinfo {author} {\bibfnamefont {H.}~\bibnamefont {Kuniki}},
  \bibinfo {author} {\bibfnamefont {H.}~\bibnamefont {Kurisu}}, \bibinfo
  {author} {\bibfnamefont {M.}~\bibnamefont {Matsuura}},\ and\ \bibinfo
  {author} {\bibfnamefont {P.}~\bibnamefont {Jang}},\ }\bibfield  {title}
  {\bibinfo {title} {Post-annealing effect of {YIG} ferrite thin-films
  epitaxially grown by reactive sputtering},\ }\href
  {https://doi.org/10.1002/pssa.200304515} {\bibfield  {journal} {\bibinfo
  {journal} {physica status solidi (a)}\ }\textbf {\bibinfo {volume} {201}},\
  \bibinfo {pages} {1810} (\bibinfo {year} {2004})}\BibitemShut {NoStop}%
\bibitem [{\citenamefont {Ma}\ \emph {et~al.}(2017)\citenamefont {Ma},
  \citenamefont {Liu}, \citenamefont {Wang},\ and\ \citenamefont
  {Wang}}]{Ma_2017}%
  \BibitemOpen
  \bibfield  {author} {\bibinfo {author} {\bibfnamefont {R.}~\bibnamefont
  {Ma}}, \bibinfo {author} {\bibfnamefont {M.}~\bibnamefont {Liu}}, \bibinfo
  {author} {\bibfnamefont {J.}~\bibnamefont {Wang}},\ and\ \bibinfo {author}
  {\bibfnamefont {H.}~\bibnamefont {Wang}},\ }\bibfield  {title} {\bibinfo
  {title} {The room temperature deposition of high-quality epitaxial yttrium
  iron garnet thin film via {RF} sputtering},\ }\href
  {https://doi.org/10.1016/j.jallcom.2017.02.275} {\bibfield  {journal}
  {\bibinfo  {journal} {Journal of Alloys and Compounds}\ }\textbf {\bibinfo
  {volume} {708}},\ \bibinfo {pages} {213} (\bibinfo {year}
  {2017})}\BibitemShut {NoStop}%
\bibitem [{\citenamefont {Ferrand}\ \emph {et~al.}(1972)\citenamefont
  {Ferrand}, \citenamefont {Daval},\ and\ \citenamefont
  {Joubert}}]{Ferrand_1972}%
  \BibitemOpen
  \bibfield  {author} {\bibinfo {author} {\bibfnamefont {B.}~\bibnamefont
  {Ferrand}}, \bibinfo {author} {\bibfnamefont {J.}~\bibnamefont {Daval}},\
  and\ \bibinfo {author} {\bibfnamefont {J.}~\bibnamefont {Joubert}},\
  }\bibfield  {title} {\bibinfo {title} {Heteroepitaxial growth of single
  crystal films of {YIG} on {GdGaG} substrates by hydrothermal synthesis},\
  }\href {https://doi.org/10.1016/0022-0248(72)90262-x} {\bibfield  {journal}
  {\bibinfo  {journal} {Journal of Crystal Growth}\ }\textbf {\bibinfo {volume}
  {17}},\ \bibinfo {pages} {312} (\bibinfo {year} {1972})}\BibitemShut
  {NoStop}%
\bibitem [{\citenamefont {Van}\ \emph {et~al.}(2018)\citenamefont {Van},
  \citenamefont {Surabhi}, \citenamefont {Dongquoc}, \citenamefont {Kuchi},
  \citenamefont {Yoon},\ and\ \citenamefont {Jeong}}]{Cao_Van_2018}%
  \BibitemOpen
  \bibfield  {author} {\bibinfo {author} {\bibfnamefont {P.~C.}\ \bibnamefont
  {Van}}, \bibinfo {author} {\bibfnamefont {S.}~\bibnamefont {Surabhi}},
  \bibinfo {author} {\bibfnamefont {V.}~\bibnamefont {Dongquoc}}, \bibinfo
  {author} {\bibfnamefont {R.}~\bibnamefont {Kuchi}}, \bibinfo {author}
  {\bibfnamefont {S.-G.}\ \bibnamefont {Yoon}},\ and\ \bibinfo {author}
  {\bibfnamefont {J.-R.}\ \bibnamefont {Jeong}},\ }\bibfield  {title} {\bibinfo
  {title} {Effect of annealing temperature on surface morphology and ultralow
  ferromagnetic resonance linewidth of yttrium iron garnet thin film grown by
  rf sputtering},\ }\href {https://doi.org/10.1016/j.apsusc.2017.11.129}
  {\bibfield  {journal} {\bibinfo  {journal} {Applied Surface Science}\
  }\textbf {\bibinfo {volume} {435}},\ \bibinfo {pages} {377} (\bibinfo {year}
  {2018})}\BibitemShut {NoStop}%
\bibitem [{\citenamefont {Gates-Rector}\ and\ \citenamefont
  {Blanton}(2019)}]{Gates_Rector_2019_YIG}%
  \BibitemOpen
  \bibfield  {author} {\bibinfo {author} {\bibfnamefont {S.}~\bibnamefont
  {Gates-Rector}}\ and\ \bibinfo {author} {\bibfnamefont {T.}~\bibnamefont
  {Blanton}},\ }\bibfield  {title} {\bibinfo {title} {The powder diffraction
  file: a quality materials characterization database - pdf card:
  00-043-0507},\ }\href {https://doi.org/10.1017/s0885715619000812} {\bibfield
  {journal} {\bibinfo  {journal} {Powder Diffraction}\ }\textbf {\bibinfo
  {volume} {34}},\ \bibinfo {pages} {352} (\bibinfo {year} {2019})}\BibitemShut
  {NoStop}%
\bibitem [{\citenamefont {Musa}\ \emph {et~al.}(2017)\citenamefont {Musa},
  \citenamefont {Azis}, \citenamefont {Osman}, \citenamefont {Hassan},\ and\
  \citenamefont {Zangina}}]{Musa_2017}%
  \BibitemOpen
  \bibfield  {author} {\bibinfo {author} {\bibfnamefont {M.~A.}\ \bibnamefont
  {Musa}}, \bibinfo {author} {\bibfnamefont {R.~S.}\ \bibnamefont {Azis}},
  \bibinfo {author} {\bibfnamefont {N.~H.}\ \bibnamefont {Osman}}, \bibinfo
  {author} {\bibfnamefont {J.}~\bibnamefont {Hassan}},\ and\ \bibinfo {author}
  {\bibfnamefont {T.}~\bibnamefont {Zangina}},\ }\bibfield  {title} {\bibinfo
  {title} {Structural and magnetic properties of yttrium iron garnet ({YIG})
  and yttrium aluminum iron garnet ({YAlG}) nanoferrite via sol-gel
  synthesis},\ }\href {https://doi.org/10.1016/j.rinp.2017.02.038} {\bibfield
  {journal} {\bibinfo  {journal} {Results in Physics}\ }\textbf {\bibinfo
  {volume} {7}},\ \bibinfo {pages} {1135} (\bibinfo {year} {2017})}\BibitemShut
  {NoStop}%
\bibitem [{\citenamefont {Patterson}(1939)}]{Patterson_1939}%
  \BibitemOpen
  \bibfield  {author} {\bibinfo {author} {\bibfnamefont {A.~L.}\ \bibnamefont
  {Patterson}},\ }\bibfield  {title} {\bibinfo {title} {The scherrer formula
  for x-ray particle size determination},\ }\href
  {https://doi.org/10.1103/physrev.56.978} {\bibfield  {journal} {\bibinfo
  {journal} {Physical Review}\ }\textbf {\bibinfo {volume} {56}},\ \bibinfo
  {pages} {978} (\bibinfo {year} {1939})}\BibitemShut {NoStop}%
\bibitem [{\citenamefont {Onbasli}\ \emph {et~al.}(2014)\citenamefont
  {Onbasli}, \citenamefont {Kehlberger}, \citenamefont {Kim}, \citenamefont
  {Jakob}, \citenamefont {Kläui}, \citenamefont {Chumak}, \citenamefont
  {Hillebrands},\ and\ \citenamefont {Ross}}]{Onbasli_2014}%
  \BibitemOpen
  \bibfield  {author} {\bibinfo {author} {\bibfnamefont {M.~C.}\ \bibnamefont
  {Onbasli}}, \bibinfo {author} {\bibfnamefont {A.}~\bibnamefont {Kehlberger}},
  \bibinfo {author} {\bibfnamefont {D.~H.}\ \bibnamefont {Kim}}, \bibinfo
  {author} {\bibfnamefont {G.}~\bibnamefont {Jakob}}, \bibinfo {author}
  {\bibfnamefont {M.}~\bibnamefont {Kläui}}, \bibinfo {author} {\bibfnamefont
  {A.~V.}\ \bibnamefont {Chumak}}, \bibinfo {author} {\bibfnamefont
  {B.}~\bibnamefont {Hillebrands}},\ and\ \bibinfo {author} {\bibfnamefont
  {C.~A.}\ \bibnamefont {Ross}},\ }\bibfield  {title} {\bibinfo {title} {Pulsed
  laser deposition of epitaxial yttrium iron garnet films with low gilbert
  damping and bulk-like magnetization},\ }\href
  {https://doi.org/10.1063/1.4896936} {\bibfield  {journal} {\bibinfo
  {journal} {{APL} Materials}\ }\textbf {\bibinfo {volume} {2}},\ \bibinfo
  {pages} {106102} (\bibinfo {year} {2014})}\BibitemShut {NoStop}%
\bibitem [{\citenamefont {CrysTech}()}]{CrysTec_YAG}%
  \BibitemOpen
  \bibfield  {author} {\bibinfo {author} {\bibnamefont {CrysTech}},\
  }\href@noop {} {\bibinfo {title} {Properties yag substrates}},\ \bibinfo
  {howpublished} {http://www.crystec.de/yag-e.html},\ \bibinfo {note}
  {accessed: 31 Oct 2020}\BibitemShut {NoStop}%
\bibitem [{\citenamefont {Gilleo}\ and\ \citenamefont
  {Geller}(1958)}]{Gilleo_1958}%
  \BibitemOpen
  \bibfield  {author} {\bibinfo {author} {\bibfnamefont {M.~A.}\ \bibnamefont
  {Gilleo}}\ and\ \bibinfo {author} {\bibfnamefont {S.}~\bibnamefont
  {Geller}},\ }\bibfield  {title} {\bibinfo {title} {Magnetic and
  crystallographic properties of substituted yttrium-iron garnet,
  $3\mathrm{Y}_2 \mathrm{O}_3 \cdotp x \mathrm{M}_2 \mathrm{O}_3 \cdotp (5-x)
  \mathrm{Fe}_2 \mathrm{O}_3$},\ }\href
  {https://doi.org/10.1103/physrev.110.73} {\bibfield  {journal} {\bibinfo
  {journal} {Physical Review}\ }\textbf {\bibinfo {volume} {110}},\ \bibinfo
  {pages} {73} (\bibinfo {year} {1958})}\BibitemShut {NoStop}%
\bibitem [{\citenamefont {Mitra}(2017)}]{Mitra_2017}%
  \BibitemOpen
  \bibfield  {author} {\bibinfo {author} {\bibfnamefont {A.}~\bibnamefont
  {Mitra}},\ }\emph {\bibinfo {title} {Structural and magnetic properties of
  YIG thin films and interfacial origin of magnetisation suppression}},\
  \href@noop {} {Ph.D. thesis},\ \bibinfo  {school} {University of Leeds}
  (\bibinfo {year} {2017})\BibitemShut {NoStop}%
\bibitem [{\citenamefont {Kittel}(1948)}]{Kittel_1948}%
  \BibitemOpen
  \bibfield  {author} {\bibinfo {author} {\bibfnamefont {C.}~\bibnamefont
  {Kittel}},\ }\bibfield  {title} {\bibinfo {title} {On the theory of
  ferromagnetic resonance absorption},\ }\href
  {https://doi.org/10.1103/physrev.73.155} {\bibfield  {journal} {\bibinfo
  {journal} {Physical Review}\ }\textbf {\bibinfo {volume} {73}},\ \bibinfo
  {pages} {155} (\bibinfo {year} {1948})}\BibitemShut {NoStop}%
\bibitem [{\citenamefont {Maier-Flaig}\ \emph {et~al.}(2018)\citenamefont
  {Maier-Flaig}, \citenamefont {Goennenwein}, \citenamefont {Ohshima},
  \citenamefont {Shiraishi}, \citenamefont {Gross}, \citenamefont {Huebl},\
  and\ \citenamefont {Weiler}}]{Maier_Flaig_2018}%
  \BibitemOpen
  \bibfield  {author} {\bibinfo {author} {\bibfnamefont {H.}~\bibnamefont
  {Maier-Flaig}}, \bibinfo {author} {\bibfnamefont {S.~T.~B.}\ \bibnamefont
  {Goennenwein}}, \bibinfo {author} {\bibfnamefont {R.}~\bibnamefont
  {Ohshima}}, \bibinfo {author} {\bibfnamefont {M.}~\bibnamefont {Shiraishi}},
  \bibinfo {author} {\bibfnamefont {R.}~\bibnamefont {Gross}}, \bibinfo
  {author} {\bibfnamefont {H.}~\bibnamefont {Huebl}},\ and\ \bibinfo {author}
  {\bibfnamefont {M.}~\bibnamefont {Weiler}},\ }\bibfield  {title} {\bibinfo
  {title} {Note: Derivative divide, a method for the analysis of broadband
  ferromagnetic resonance in the frequency domain},\ }\href
  {https://doi.org/10.1063/1.5045135} {\bibfield  {journal} {\bibinfo
  {journal} {Review of Scientific Instruments}\ }\textbf {\bibinfo {volume}
  {89}},\ \bibinfo {pages} {076101} (\bibinfo {year} {2018})}\BibitemShut
  {NoStop}%
\bibitem [{\citenamefont {Sun}\ \emph {et~al.}(2012)\citenamefont {Sun},
  \citenamefont {Song}, \citenamefont {Chang}, \citenamefont {Kabatek},
  \citenamefont {Jantz}, \citenamefont {Schneider}, \citenamefont {Wu},
  \citenamefont {Schultheiss},\ and\ \citenamefont {Hoffmann}}]{Sun_2012}%
  \BibitemOpen
  \bibfield  {author} {\bibinfo {author} {\bibfnamefont {Y.}~\bibnamefont
  {Sun}}, \bibinfo {author} {\bibfnamefont {Y.-Y.}\ \bibnamefont {Song}},
  \bibinfo {author} {\bibfnamefont {H.}~\bibnamefont {Chang}}, \bibinfo
  {author} {\bibfnamefont {M.}~\bibnamefont {Kabatek}}, \bibinfo {author}
  {\bibfnamefont {M.}~\bibnamefont {Jantz}}, \bibinfo {author} {\bibfnamefont
  {W.}~\bibnamefont {Schneider}}, \bibinfo {author} {\bibfnamefont
  {M.}~\bibnamefont {Wu}}, \bibinfo {author} {\bibfnamefont {H.}~\bibnamefont
  {Schultheiss}},\ and\ \bibinfo {author} {\bibfnamefont {A.}~\bibnamefont
  {Hoffmann}},\ }\bibfield  {title} {\bibinfo {title} {Growth and ferromagnetic
  resonance properties of nanometer-thick yttrium iron garnet films},\ }\href
  {https://doi.org/10.1063/1.4759039} {\bibfield  {journal} {\bibinfo
  {journal} {Applied Physics Letters}\ }\textbf {\bibinfo {volume} {101}},\
  \bibinfo {pages} {152405} (\bibinfo {year} {2012})}\BibitemShut {NoStop}%
\bibitem [{\citenamefont {Liu}\ \emph {et~al.}(2014)\citenamefont {Liu},
  \citenamefont {Chang}, \citenamefont {Vlaminck}, \citenamefont {Sun},
  \citenamefont {Kabatek}, \citenamefont {Hoffmann}, \citenamefont {Deng},\
  and\ \citenamefont {Wu}}]{Liu_2014}%
  \BibitemOpen
  \bibfield  {author} {\bibinfo {author} {\bibfnamefont {T.}~\bibnamefont
  {Liu}}, \bibinfo {author} {\bibfnamefont {H.}~\bibnamefont {Chang}}, \bibinfo
  {author} {\bibfnamefont {V.}~\bibnamefont {Vlaminck}}, \bibinfo {author}
  {\bibfnamefont {Y.}~\bibnamefont {Sun}}, \bibinfo {author} {\bibfnamefont
  {M.}~\bibnamefont {Kabatek}}, \bibinfo {author} {\bibfnamefont
  {A.}~\bibnamefont {Hoffmann}}, \bibinfo {author} {\bibfnamefont
  {L.}~\bibnamefont {Deng}},\ and\ \bibinfo {author} {\bibfnamefont
  {M.}~\bibnamefont {Wu}},\ }\bibfield  {title} {\bibinfo {title}
  {Ferromagnetic resonance of sputtered yttrium iron garnet nanometer films},\
  }\href {https://doi.org/10.1063/1.4852135} {\bibfield  {journal} {\bibinfo
  {journal} {Journal of Applied Physics}\ }\textbf {\bibinfo {volume} {115}},\
  \bibinfo {pages} {17A501} (\bibinfo {year} {2014})}\BibitemShut {NoStop}%
\bibitem [{\citenamefont {Coey}(2001)}]{Coey_2001}%
  \BibitemOpen
  \bibfield  {author} {\bibinfo {author} {\bibfnamefont {J.~M.~D.}\
  \bibnamefont {Coey}},\ }\href {https://doi.org/10.1017/cbo9780511845000}
  {\emph {\bibinfo {title} {Magnetism and Magnetic Materials}}}\ (\bibinfo
  {publisher} {Cambridge University Press},\ \bibinfo {year}
  {2001})\BibitemShut {NoStop}%
\bibitem [{\citenamefont {Shen}\ \emph {et~al.}(2009)\citenamefont {Shen},
  \citenamefont {Xu}, \citenamefont {Wu}, \citenamefont {Zhao},\ and\
  \citenamefont {Shi}}]{Shen_2009}%
  \BibitemOpen
  \bibfield  {author} {\bibinfo {author} {\bibfnamefont {H.}~\bibnamefont
  {Shen}}, \bibinfo {author} {\bibfnamefont {J.}~\bibnamefont {Xu}}, \bibinfo
  {author} {\bibfnamefont {A.}~\bibnamefont {Wu}}, \bibinfo {author}
  {\bibfnamefont {J.}~\bibnamefont {Zhao}},\ and\ \bibinfo {author}
  {\bibfnamefont {M.}~\bibnamefont {Shi}},\ }\bibfield  {title} {\bibinfo
  {title} {Magnetic and thermal properties of perovskite {Y}{F}e{O}$_3$ single
  crystals},\ }\href@noop {} {\bibfield  {journal} {\bibinfo  {journal}
  {Materials Science and Engineering: B}\ }\textbf {\bibinfo {volume} {157}},\
  \bibinfo {pages} {77} (\bibinfo {year} {2009})}\BibitemShut {NoStop}%
\bibitem [{\citenamefont {Rosales-González}\ \emph {et~al.}(2019)\citenamefont
  {Rosales-González}, \citenamefont {Sánchez-De~Jesús}, \citenamefont
  {Pedro-García}, \citenamefont {Cortés-Escobedo}, \citenamefont
  {Ramírez-Cardona},\ and\ \citenamefont {Bolarín-Miró}}]{Rosales_2019}%
  \BibitemOpen
  \bibfield  {author} {\bibinfo {author} {\bibfnamefont {O.}~\bibnamefont
  {Rosales-González}}, \bibinfo {author} {\bibfnamefont {F.}~\bibnamefont
  {Sánchez-De~Jesús}}, \bibinfo {author} {\bibfnamefont {F.}~\bibnamefont
  {Pedro-García}}, \bibinfo {author} {\bibfnamefont {C.~A.}\ \bibnamefont
  {Cortés-Escobedo}}, \bibinfo {author} {\bibfnamefont {M.}~\bibnamefont
  {Ramírez-Cardona}},\ and\ \bibinfo {author} {\bibfnamefont {A.~M.}\
  \bibnamefont {Bolarín-Miró}},\ }\bibfield  {title} {\bibinfo {title}
  {Enhanced multiferroic properties of {Y}{F}e{O}$_3$ by doping with
  {B}i$^{3+}$},\ }\href {https://www.mdpi.com/1996-1944/12/13/2054} {\bibfield
  {journal} {\bibinfo  {journal} {Materials}\ }\textbf {\bibinfo {volume} {12}}
  (\bibinfo {year} {2019})}\BibitemShut {NoStop}%
\bibitem [{\citenamefont {Dietrich}\ \emph {et~al.}(2008)\citenamefont
  {Dietrich}, \citenamefont {Hertel}, \citenamefont {Huber}, \citenamefont
  {Weiss}, \citenamefont {Schäfer},\ and\ \citenamefont
  {Zweck}}]{Dietrich_2008}%
  \BibitemOpen
  \bibfield  {author} {\bibinfo {author} {\bibfnamefont {C.}~\bibnamefont
  {Dietrich}}, \bibinfo {author} {\bibfnamefont {R.}~\bibnamefont {Hertel}},
  \bibinfo {author} {\bibfnamefont {M.}~\bibnamefont {Huber}}, \bibinfo
  {author} {\bibfnamefont {D.}~\bibnamefont {Weiss}}, \bibinfo {author}
  {\bibfnamefont {R.}~\bibnamefont {Schäfer}},\ and\ \bibinfo {author}
  {\bibfnamefont {J.}~\bibnamefont {Zweck}},\ }\bibfield  {title} {\bibinfo
  {title} {Influence of perpendicular magnetic fields on the domain structure
  of permalloy microstructures grown on thin membranes},\ }\href@noop {}
  {\bibfield  {journal} {\bibinfo  {journal} {Physical Review B}\ }\textbf
  {\bibinfo {volume} {77}} (\bibinfo {year} {2008})}\BibitemShut {NoStop}%
\bibitem [{\citenamefont {Sloika}\ \emph {et~al.}(2014)\citenamefont {Sloika},
  \citenamefont {Kravchuk}, \citenamefont {Sheka},\ and\ \citenamefont
  {Gaididei}}]{Sloika_2014}%
  \BibitemOpen
  \bibfield  {author} {\bibinfo {author} {\bibfnamefont {M.~I.}\ \bibnamefont
  {Sloika}}, \bibinfo {author} {\bibfnamefont {V.~P.}\ \bibnamefont
  {Kravchuk}}, \bibinfo {author} {\bibfnamefont {D.~D.}\ \bibnamefont
  {Sheka}},\ and\ \bibinfo {author} {\bibfnamefont {Y.}~\bibnamefont
  {Gaididei}},\ }\bibfield  {title} {\bibinfo {title} {Curvature induced
  chirality symmetry breaking in vortex core switching phenomena},\ }\href
  {https://doi.org/10.1063/1.4884957} {\bibfield  {journal} {\bibinfo
  {journal} {Applied Physics Letters}\ }\textbf {\bibinfo {volume} {104}},\
  \bibinfo {pages} {252403} (\bibinfo {year} {2014})}\BibitemShut {NoStop}%
\end{thebibliography}%

\end{document}